\newif\iffigs
  \figsfalse
\documentclass[12pt]{article}
\usepackage{colordvi}
\usepackage{amsmath}
\usepackage{latexsym,amssymb}
\newsavebox{\uuunit}
\sbox{\uuunit}
                             {\setlength{\unitlength}{0.825em}
                                      \begin{picture}(0.6,0.7)
                                                                  \thinlines
                                                                  \put(0,0){\line(1,0){0.5}}
                                                                  \put(0.15,0){\line(0,1){0.7}}
                                                                  \put(0.35,0){\line(0,1){0.8}}
                                                                 \multiput(0.3,0.8)(-0.04,-0.02){10}{\rule{0.5pt}{0.5pt}}
                                      \end {picture}}

\makeatletter \@addtoreset{equation}{section} \makeatother

\newcommand{\E}{\mathop{\rm E}}
\newcommand{\SU}{\mathop{\rm SU}}
\newcommand{\PSU}{\mathop{\rm PSU}}
\newcommand{\SO}{\mathop{\rm SO}}

\newcommand{\U}{\mathop{\rm {}U}}
\newcommand{\USp}{\mathop{\rm {}USp}}
\newcommand{\Sp}{\mathop{\rm {}Sp}}
\newcommand{\OSp}{\mathop{\rm {}OSp}}

\newcommand{\SL}{\mathop{\rm SL}}

\newcommand{\cN}{\mathcal{N}}
\newcommand{\cP}{\mathcal{P}}
\newcommand{\sM}{{\Scr M}}

\newcommand{\eq}[1]{(\ref{#1})}

\def\a{\alpha}
\def\b{\beta}
\def\g{\gamma}

\def\oN{\buildrel{\circ}\over N}

\def\bfone{\relax{\rm 1\kern-.35em 1}}
\DeclareFontFamily{U}{rsf}{} \DeclareFontShape{U}{rsf}{m}{n}{
  <5> <6> rsfs5 <7> <8> <9> rsfs7 <10-> rsfs10}{}
\DeclareMathAlphabet\Scr{U}{rsf}{m}{n}

\begin{document}
\begin{titlepage}
\begin{flushright}
CERN-PH-TH/2008-206\\
UCB-PTH-08/69\\
DISTA-2008
\end{flushright}
\vskip 5mm

  \begin{center}{\LARGE \bf  Exceptional $\cN=6$ and $\cN=2$ \\
  $AdS_4$ Supergravity, \\ \vskip .2cm
  and
  Zero-Center Modules}
\vskip 1.5cm
{L. Andrianopoli$^{1}$,  R. D'Auria$^1$, S. Ferrara$^2$, P. A. Grassi$^3$ and M. Trigiante$^1$}   \end{center}
 \vskip 3mm
\begin{center}
 \noindent
{\small $^1$ Dipartimento di Fisica,
  Politecnico di Torino, Corso Duca degli Abruzzi 24, \\
$~~~$ I-10129
  Turin, Italy and INFN, Sezione di Torino, Italy.
 }

\noindent
{\small
$^2$
Physics Department,Theory Unit, CERN,
CH 1211, Geneva 23, Switzerland,\\
$~~~$  Miller Institute for Basic Research in Science, University of California, Berkeley \\
$~~~$ and INFN - Laboratori Nazionali di Frascati,
00044 Frascati, Italy.
 }

\noindent
 {\small
$^3$
DISTA, University of Eastern Piedmont,
via Bellini 25/g, I-15100 Alessandria, \\$~~$  Italy
and INFN, Sezione di Torino, Italy.
 }
\end{center}


\vfill
\begin{abstract}
{\small We study the gauging of the orthosymplectic algebras
$\OSp(6|4)\times \SO(2)$ and its ``dual" $\OSp(2|4)\times \SO(6)$,
both based on supergravities with the same exceptional coset
$\SO^*(12)/\U(6)$,  and
 gauge group $\SO(6)\times \SO(2)$. The two dual theories are
obtained by two different truncations of gauged $\cN=8$ $AdS_4$
supergravity. We explicitly study the gauge sector of the two dual
theories with the most general group allowed by supersymmetry. In
the ungauged (super-Poincar\'e) case they  exhibit the same
(large) black-hole attractor solutions with dual relations between
the $1/\cN$-BPS and non-BPS configurations. The $\cN=6$ gravity
multiplet has also the exceptional property to be a {\em
zero-center module} of $\OSp(6|4)$, as it is the case for
superconformal Yang--Mills theory in four dimensions based on
$\SU(2,2|n)$ ($\PSU(2,2|4)$ for $n=4$) or $\OSp(n|4)$.}

\end{abstract}

\vskip 3mm
 \footnotesize{

 \texttt{laura.andrianopoli@polito.it};

 \texttt{riccardo.dauria@polito.it};

 \texttt{pgrassi@cern.ch};

  \texttt{sergio.ferrara@cern.ch};

   \texttt{mario.trigiante@polito.it}.}

\end{titlepage} \vfill \eject
\section{Introduction} \label{intro}
Gauged supergravities pertain to a  topical subject of
investigation because they are related to the possibility of
turning on a scalar potential in an effective theory of gravity
which can stabilize many of the scalar modes of the theory.
Popular examples of such gaugings are those obtained by flux vacua
in superstring theory \cite{flux}.

Particular classes of these vacua can show residual supersymmetry both in Minkowski
or anti de Sitter space, depending on the nature of the gauging of a given theory.

Minkowski vacua with residual supersymmetry correspond to theories with
 $\cN$-extended Poincar\'e supersymmetry, with $0\leq \cN <8$ at $D=4$.
Typical compactifications giving rise to such vacua are those based on
 {\em generalized Calabi--Yau manifolds} \cite{gcy} or on {\em twisted tori} \cite{twisted},
  the latter being the modern version of the gauging of {\em flat groups} \`a
  la Scherk--Schwarz \cite{ss}.
 These vacua give a realization of the so called no-scale models as they usually provide
 (partial) supersymmetry breaking with sliding gravitino mass and zero vacuum energy.
 In these compactifications  one can then turn on further fluxes such as those
 giving rise to black holes and study interesting phenomena such as the
  {\em attractor mechanism} \cite{attractor}.

 Another class of flux compactifications, whose interest is further
 motivated by additional physical properties, is the one corresponding to anti de Sitter vacua.
 These vacua are related to the famous $AdS_{d+1}/CFT_d$ correspondence,
 the most popular one being the $d=4$ case \cite{ads/cft,ferrarads}. In this case the supergravity in
 question is the maximally extended gauged supergravity at $D=5$
 based on the superalgebra $\SU(2,2|4)$ \cite{Gunaydin:1985cu}.
 Other examples of anti de Sitter supergravities  relevant for the
 $AdS_{d+1}/CFT_d$ correspondence are those at $d=3$ and $d=6$, based
 on two different real forms of the orthosymplectic algebra $\OSp(8|4)$.
 However in recent times other classes of $AdS/CFT$ dual theories have been found,
 after realizing  that superconformal invariant Chern--Simons theories can be
 constructed whose dual bulk supergravity theories correspond to lower $\cN$
 orthosymplectic algebras $\OSp(\cN|4)$, with $2\leq \cN\leq 6$
  \cite{Bagger:2006sk,Gustavsson:2007vu,Bagger:2007vi,Bagger:2007jr,Aharony:2008ug,Benna:2008zy}.

 It is the aim of the present paper to investigate some of the exceptional properties of the $\cN=6$ gauged supergravity theory and its
 ``dual relation" to an $\cN=2$ theory based on the {\em exceptional model} related to $J_3^H$, one of the four degree-three
 Jordan algebras of the magic square introduced in
 \cite{freud,Gunaydin:1983rk,Gunaydin:1983bi}.
Already in the ungauged case  $\cN=6$ and  $\cN=2$ supergravity,
based on symmetric scalar manifold $\SO^*(12)/\U(6)$, exhibit a
duality relation, since, although different in the fermionic
sector, they have the same bosonic content. In particular they
exhibit the same (large ) extremal black-hole attractor solutions
where the role of the $BPS$ and non $BPS$ configurations in the
two theories are exchanged. The superstring origin of these two
ungauged theories was investigated in \cite{Dolivet:2007sz} in the
context of compactifications on asymmetric orbifolds. Let us
remark, moreover, that  the duality between the  $\cN=6$ and
$\cN=2$ four-dimensional theories has a three-dimensional
counterpart in the duality between $\cN=12$ supergravity and the
$\cN=4$ theory based on the exceptional quaternionic manifold
$\E_{7(-5)}/[\SU(2)\times\SO(12)]$ \cite{Cecotti:1988qn,de
Wit:1992up}.

 In the present investigation we concentrate on the gauging of these theories and
 we will show that both
 of these models can be obtained as truncations of the gauged $\cN=8$
 theory of \cite{de Wit:1982ig}, with gauge structure $\OSp(6|4)\times \SO(2)$ in the $\cN=6$ case,
 and $\OSp(2|4)\times \SO(6)$ in the $\cN=2$ case.
 These superalgebras  are indeed both subalgebras of the $\OSp(8|4)$ superalgebra,
 when one retains  respectively  24 or 8 of the  original 32  fermionic generators
 (anti de Sitter spinors).
As far as the gauging is concerned, we analyze  the consistency of
the truncation procedure and, by use of the embedding tensor
formalism, we give a detailed analysis of the gauge sector of both
theories, for a generic group. We then work out the details in the
particular case of the $\SO(2)\times \SO(6)$ gaugings, and determine
the explicit form of the fermionic shifts and the scalar potential.

 From a four-dimensional point of view, the $\cN=6$ theory is obtained just
 by gauging the $\SO(6)$ gauge group inside the U-duality group $\SO^*(12)$,
 which is the maximal group commuting with $\SU(2)$ inside $\E_{7(7)}$ (the
 U-duality group of $\cN=8$, $D=4$ supergravity).
On the other hand, the $\cN=2$ theory is obtained by gauging both
$\SO(6) \subset\SO^*(12)$ and $\SO(2)\subset \SU(2)$, the global R-symmetry
of the truncated $\cN=2$ theory, which does not participate to the gauging in the $\cN=6$ case.

Note that in the $\cN=6$ theory the spectrum is obtained from the
$\cN=8$ spectrum by projecting out all $\SU(2)$ non-singlets. An
extra $\SO(2)$ abelian symmetry, implied by the structure of the
$\cN=6$ supergravity multiplet remains, commuting with the
superalgebra. On the other hand in the $\cN=2$ theory the $\SO(6)$
symmetry commutes with the supercharges and it then merely acts as
a matter flavor symmetry. It can therefore be gauged by the
$\cN=2$ matter vectors, which precisely sit in the adjoint
representation of $\SO(6)$. In fact, if in the truncation we would
only keep $\SO(6)$ singlets, we would obtain pure $\cN=2$ anti de
Sitter supergravity, with the gravitino mass induced by a $\cN=2$
Fayet--Iliopoulos term. This way to generate the gravitino mass is
indeed common to all $\cN=2$ theories in $AdS_4$, when
hypermultiplets are absent. The $\cN=2$ theory under investigation
is a particular case of $AdS_4$ theories with superalgebras
$\OSp(2/4)\times G_e$, where $G_e$ denotes the gauge symmetry,
corresponding to the gauged isometries of the vector multiplets
scalar manifold. Since  the manifold spanned by the scalars in
$\cN=2$ theory is only constrained by supersymmetry to be
special-K\"ahler, it can be chosen to have any isometry  $G$, so
that we can always accommodate $G_e \subset G$ such that $G_e$ is
an {\em electric subgroup} of $G$.
    The simplest example is the minimal series \cite{Luciani:1977hp,deWit:1984pk}
    where $G=\SU(1,m)$ and $G_e$ can be embedded in $G$ for $m\geq \mathrm{Adj}\,[G_e]$.
    In this case the gauge symmetry is $\SO(2)\times G_e$, with an $AdS_4$ vacuum
    which is $\SO(2)\times G_e$ invariant. A similar embedding of $G_e$ in $G$ can
    be given also for $\cN=3$ supergravity with gauge group $\SO(3)\times G_e$ and
    for $\cN=5$ with gauge group $\SO(5)$. The latter theory was obtained long ago
    by de Wit and Nicolai \cite{deWit:1981yv}.

    Recently many of these theories have been shown to have a $CFT_3$ dual as a
    Chern--Simons gauge theory. In this case the $G_e$ commuting with $\OSp(\cN|4)$
    is identified with some flavor symmetry of the conformal theory matter multiplets.
The $AdS_4$ theory with lowest supersymmetry is based on the
$\OSp(1|4)$ superalgebra, with no gauge sector. In this case,
symmetric $AdS_4$ vacua with any gauge group can be accommodated.

Note that the $\cN=6$ case has a special role among the $\cN$-extended theories,
because it is the only one  with $\cN>4$ which contains an additional $\U(1)$
conserved current, and further  because it is the only one which has a zero-center
module supergravity multiplet, unlike $\cN\neq 6$
 orthosymplectic supergravities \cite{Flato:1984du}.

The paper is organized as follows: In Section \ref{antonio} we
point out some exceptional properties of the $\OSp(6|4)$ algebra:
to have the gravity multiplet as a {\em zero center module},
according to Flato and Fronsdal \cite{Flato:1984du}, and to have a
zero Killing--Cartan form, which makes it more similar to the
$\SU(2,2|4)$ case than other orthosymplectic cases. The definition
of such exceptional properties and of their possible physical
implications is reported in this section. In Section \ref{n6n2} we
discuss the $\cN=6$ and $\cN=2$ dual theories, both at the
ungauged and gauged level, as they come from different truncations
of $\cN=8$ (anti de Sitter or Poincar\'e) supergravity in four
dimensions and we make some comments on the relation of $\cN=6$
supergravity with its ancestor theory, namely IIA supergravity
compactified on $AdS_4\times\mathbb{C}P_3$
\cite{Nilsson:1984bj,Arutyunov:2008if,Fre:2008qr,Stefanski:2008ik,D'Auria:2008cw},
which is the higher dimensional theory underlying $\cN=6$
supergravity. In Section \ref{nove} we briefly discuss a different
$\cN=2$ truncation of the $\cN=8$ theory, in which the
supergravity multiplet is coupled to 10 hypermultiplets and no
vector multiplet. In Appendix \ref{vcf} examples of supergroups
and supercosets with vanishing Killing-Cartan form are given. In
Appendix \ref{decs} the reader may find a list of branchings and
decompositions which are used in our analysis. Finally in Appendix
\ref{masst} the spin-$1/2$ mass terms in the $\cN=6$ and $\cN=2$
 theories are given.

\section{Zero center modules}\label{antonio}

From a group-theoretical point of view, $\cN=6$ supergravity on
$AdS_4$ has two reasons for being exceptional: {\it 1}) the
superalgebra on which is based, namely $\OSp(6|4)$ has zero
Killing-Cartan form and {\it 2}) the zero-center module coincides
with the supergravity multiplet. In the present section, we recall
some basic facts about orthosymplectic superalgebras, the relation
with supergravity backgrounds, the Killing-Cartan forms and the
zero-center modules.

The compactification on $AdS_4 \times {\mathbb CP}_3$ of 10d type
IIA string theory can be completely discussed in terms of the
supermanifold
\begin{equation}\label{suA}
\frac{\OSp(6|4)}{ \U(3) \times \SO(1,3)}\,.
\end{equation}
The bosonic subgroup of the isometry group $\OSp(6|4)$ is $\SO(6)
\times \Sp(4)$ and therefore the bosonic coset $\SO(6) \times
\Sp(4)/ \U(3) \times \SO(1,3)$ is the direct product of the
homogeneous spaces ${\mathbb CP}^3 \times AdS_4$. In addition,
there are fluxes associated to $F^{(4)} = g \epsilon$ and $F^{(2)}
= k {\cal J}$ where $\epsilon$ and ${\cal J}$ are the Levi-Civita
tensor in $AdS_4$ and the K\"alher form on ${\mathbb CP}^3$,
respectively. The fluxes $g$ and $k$ appear in the commutation
relations of the supercharges. This background is a solution of
type IIA supergravity
 in 10d (see
 \cite{Nilsson:1984bj,Arutyunov:2008if,Fre:2008qr,D'Auria:2008cw}).
The fermionic sector is indeed described by 24 anticommuting supecharges $Q^A_\a$ in
the fundamental representations of $\SO(6)$ and $\Sp(4)$. The superalgebra associated
to (\ref{suA}) is given in terms of the bosonic
generators $P_{\a\b}$ (where $\a,\b=1,\dots,4$ and are the $\Sp(4)$ generators)
and $T^{AB}$ (where $A,B=1,\dots,6$ and they are $\SO(6)$ generators) and
in terms of the fermionic generators $Q^A_\a$
\begin{eqnarray}\label{suB}
&&\{Q^A_\a, Q^B_\b\} = \eta^{AB} P_{\a\b} + T^{AB} \epsilon_{\a\b}\,,  \nonumber \\\
&&[P_{\a\b}, P_{\g\delta}] = \frac{1}{2} \Big(
\epsilon_{\g(\a} P_{\b)\delta} + \epsilon_{\delta(\a} P_{\b)\g}\Big)\,, \nonumber \\
&&[T^{AB}, T^{CD}] =  \frac{1}{2} \Big(\eta^{C[A} T^{B]D} - \eta^{D[A} T^{B]C} \Big) \,, \\
&&[T^{AB}, Q^C_\a] = \eta^{C[A} Q^{B]}_\a\,,  \nonumber \ \\
&&[P_{\a\b}, Q^C_\g] = \epsilon_{\g(\a} Q^{C}_{\b)}\,.  \nonumber \
\end{eqnarray}
In order to see the presence of the constants $g$ and $k$, we
decompose the generators $P_{\a\b} = \g^m_{\a\b} P_m + g^{-1} \,
\g^{mn}_{\a\b} L_{mn}$ where $P_m$ are the generators of the coset
and $L_{mn}$ are the $\SO(1,3)$ generators, and $T^{AB} =
f^{AB}_{IJ} T^{IJ} + f^{AB}_{\bar I\bar J} T^{\bar I\bar J} +
k^{-1}\, f^{AB}_{I\bar J} T^{I\bar J}$ where $T^{IJ}, T^{\bar
I\bar J}$ are the generators of the coset $\SU(4)/\U(3)$ and
$T^{I\bar J}$ are the generators of the subgroup. Therefore, when
the algebra is decomposed into the generators of the subgroup
$\U(3) \times \SO(1,3)$ one can see the two constants
$g^{-1},k^{-1}$ multiplying the generators of the subgroup.
Accordingly, in the Maurer-Cartan equations of the coset, the
coupling constants $g$ and $k$ multiply the $H$-connections.

The form of the superalgebra is the same for any $R$-symmetry group $\SO(\cN)$.
The Killing-Cartan form for $\cN=6$ vanishes. We have to recall that the Killing-Cartan
form is defined as follows
\begin{equation}\label{suD}
K(X,Y) = \frac{1}{2}{\rm Str} \Big( {\rm ad}_X {\rm ad}_Y \Big)\,,
\end{equation}
where $X,$ are generators of the supergroup and ${\rm Str}$ is the
supertrace. In the appendix \ref{vcf}, the supergroups with
vanishing Killing-Cartan form are listed \cite{Frappat:1996pb}. On
the other hand the representations are classified according to the
invariant tensors on the Lie superalgebra denoted Casimir
operators. Given a non-degenerate Killing-Cartan metric there is a
simple way to construct the basic quadratic Casimir. However, in
general one can construct it as follows: consider the following
restricted metric (which coincides with the Killing-Cartan form on
the subgroups $\SO(6)$ and $\Sp(4)$ and on the supergenerators)
\begin{equation}\label{suC}
\langle P_{\a\b}, P_{\g\delta} \rangle = \epsilon_{\a(\g} \epsilon_{\delta)\b}\,,
\quad
\langle T^{AB}, T^{CD} \rangle = \eta^{A[C} \eta^{D]B} \,,
\quad
\langle Q^A_\a, Q^B_\b \rangle = \epsilon_{\a\b} \eta^{AB}\,,
\end{equation}
where $\langle , \rangle$ denotes the trace, and define
\begin{equation}\label{suD'}
C_2 = \epsilon^{\a\b} \epsilon^{\g\delta} \, P_{\a\g} P_{\b\delta} - \eta_{AC} \eta_{BD} \, T^{AB} T^{CD} +
\epsilon^{\a\b} \eta_{AB} \, Q_{\a}^A Q_\b^B\,.
\end{equation}
$C_2$ is constructed in terms of quadratic invariants of
$\Sp(4)\times \SO(6)$. The coefficients of the linear combination
$\OSp(6|4)$ invariant can be found by commuting $C_2$ with all the
fermionic  generators of the supergroup. \footnote{We would like
to stress the analogy with abelian Lie algebras: the
Killing-Cartan form is vanishing, but one can define an invariant
bilinear form.} For $\OSp(\cN|4)$ there are other invariant
Casimir operators that can be constructed with higher powers of
generators.

The irreducible, positive energy representations of $\Sp(4)$ are
fully characterized by the lowest value $E_0$ of the energy and by
the spin $s$ and they are denoted by $D(E_0,s)$. The massless
representations are $D(s+1,s)$ and the Dirac singleton are
$D(1/2,0)$ and $D(1,1/2)$. Among the massless representations,
$D(2,1)$ has both Casimir operators equal to zero. (The same is
also valid for the conformal group in 4d, namely $\SO(4,2)$, whose
representations $D(2,1,0)$ and $D(2,0,1)$ have vanishing Casimir
operators). Those representations are referred to as {\it
zero-center module} since the center of the enveloping algebra is
zero.  In analogy with the conformal group in 3d $\Sp(4)$ and with
$\SO(4,2)$, the zero-center module of a superalgebra is a
representation characterized by the vanishing of all super-Casimir
operators. A zero-center module is a special short representation
of a superalgebra and it plays a role similar to the vacuum state.

According to \cite{Flato:1984du}, in the case of $AdS_4$ algebras one can find
the following zero-center modules
\begin{eqnarray}
\OSp(6|4) & \cN=6 & D(3,2|1) \oplus D(5/2, 3/2|6) \oplus D(2,1|15)
\oplus D(2,1|1) \nonumber \cr &&\oplus D(3/2,1/2|20) \oplus
D(3/2,1/2|\bar 6) \oplus D(1, 0|15) \oplus D(1,0|\overline{15})\,,
  \nonumber \\
\OSp(5|4) & \cN=5 & D(5/2,3/2|1) \oplus D(2,1|5) \oplus D(2,1|1)
\oplus D(3/2,1/2|10) \nonumber \cr &&\oplus D(3/2,1/2|\bar 5)
\oplus D(1,0|\overline{10}) \oplus D(1,0|{10})
  \nonumber \\
\OSp(4|4) & \cN=4 & D(2,1|1) \oplus D(3/2,1/2|4)  \oplus D(1,0|6)
\,,
  \nonumber \\
\OSp(3|4) & \cN=3 & D(2,1|1) \oplus D(3/2,1/2|3) \oplus
D(3/2,1/2|1) \oplus D(1,0|{6}) \,,
  \nonumber \\
\OSp(2|4) & \cN=2 & D(2,1|1) \oplus D(3/2,1/2|2) \oplus
D(1,0|{2})\,.
  \nonumber \\
\OSp(1|4) & \cN=1 & D(2,1|1) \oplus D(3/2,1/2|1) \,.
\end{eqnarray}
where we have denoted by $D(s+1,s|n)$ respectively the $\Sp(4)$
representation and the dimension of the representation of the
orthogonal group $\SO(\cN)$. Notice that only $\OSp(6|4)$ has
 the supergravity multiplet
(starting with the supergravity state $D(3,2|1)$ as the zero-center
module (by the way, it is also the only supergroup of the $\OSp(\cN|4)$ with vanishing Killing-Cartan form).
 For the supergroup $\OSp(5|4)$, the zero center module is
represented by the gravitino multiplet $D(5/2, 3/2|1)$. The other
four examples have, as zero-center module, the SYM multiplet  with
$\cN=1,2,3,4$ supersymmetries. The technique to establish the
existence of unitary zero-center module representations is that of
the ``induced representations'' and it amounts to check if in the
induced representation there is the trivial representation (the
``vacuum"). In that case the module is a zero-center module.

Let us look to other series of supergroups with analogous
peculiarities. As we can read from the Appendix \ref{vcf} there
are other interesting supergroups with vanishing Killing-form
which play an important role in superstring. \footnote{Another
example is $\OSp(4|2)$, which might play a role in non-critical
strings. It has zero Killing-Cartan form and it would be
interesting to study its zero-center modules.} The most relevant
one is the case of $\PSU(2,2|4)$, with supercoset
\begin{equation}\label{suF}
\frac{\PSU(2,2|4)}{\SO(1,4) \times \SO(5)}
\end{equation}
whose bosonic part is described by $AdS_5 \times S^5$. Again, one
can study the sequence of supergroups $\SU(2,2|\cN)$ where
$\cN=1,2,3$ and for each of them  identifying the zero-center module. The
fermionic sector is described by complex supercharges $Q_I^a,
\overline Q^I_a$ (where $I$ is the $\SU(2,2)$ index and
$a=1,\dots,\cN$). However, we can observe the following fact: we can
relate the supergroup $\PSU(2,2|4)$ to the orthosymplectic
$\OSp(4|4)$ by imposing the reality condition
\cite{Berkovits:2007zk,Berkovits:2007rj}
\begin{equation}
Q_I^a = \epsilon_{IJ} \delta^{ab} \overline Q^J_b\,.
\end{equation}
The invariant tensor $\epsilon_{IJ}$ breaks the group $\SU(2,2)$ to $\SO(2,3) \sim \Sp(4)$ while
the invariant tensor $\delta^{ab}$ breaks the group $\U(n)$ down to $\SO(n)$. Therefore, we can
relate the supergroup $\PSU(2,2|4)$ with $\OSp(4|4)$ and the latter has the vector multiplet as zero-center module.

Another interesting example is the superalgebra $\SU(2,2|3)$ which
underlies the $N=6$ supergravity on $AdS_5$ with gauge group
$\U(3)$. It has a zero-center module which is the supersingleton
of $\SU(2,2|3)$. In the same way as above we can break
$\SU(2,2|3)$ down to $\OSp(3|4)$ which is the $\cN=3$ vector
multiplet in $AdS_4$ (using the topological string model
constructed on Grassmannian spaces (see
\cite{Berkovits:2008qc,Bonelli:2008us}) it should be possible to
justify the selection rules discussed in \cite{Ferrara:1998zt}).

Notice that the $\OSp(3|4)$ has the vector representation as a
zero-center module and therefore, one can argue that the
zero-center module representation of $\OSp$ type are related to
zero-center module representation of $SU$-type. To support this
argument, we notice that the case $\OSp(1|4)$ which has the
zero-center module which contains the vector multiplet, can be
obtained by reducing it from $\SU(2,2|1)$ which indeed has a
zero-center module. Indeed, one can verify that the zero-center
modules of $\SU(2,2|\cN)$ are mapped into zero-center modules of
$\OSp(\cN|4)$.


\section{Universal supergravity relations} \label{universal}
We recall that in any supergravity theory there is a universal
relation between the anti de Sitter cosmological constant and the
gravitino mass. Indeed,  for every four dimensional extended
theory supersymmetry implies that the following Ward identity
holds:
\begin{eqnarray}
\delta^A_B \,V(\phi) &=& -12\,g^2 \,\bar S^{AC} S_{CB} \,+
\,g^2\,\bar N_I^A N^I_B\label{ward}
\end{eqnarray}
where $S_{AB}$ and $N^A_I$ are scalar field dependent matrices
also appearing in the Lagrangian, the former defining the
gravitino mass-like term:
\begin{eqnarray}
2\,g\,S_{AB} \,\bar \psi^A_\mu \,\gamma^{\mu\nu}\psi^B_\nu \,+
h.c.\,,\label{psim}
\end{eqnarray}
$2\,g\,S_{AB}$ being the gravitino mass matrix, the latter
entering the spin-$1/2$ -- gravitino couplings:
\begin{eqnarray}
{\rm i}\,g \, \left(N_I^A \,\bar \lambda^I\,\gamma^{\mu}\psi_{\mu
A} \, +h.c.\right)\,,\label{chim}
\end{eqnarray}
as reviewed in \cite{D'Auria:2001kv}. Here $A,B,\cdots$ are
indices of the fundamental representation of the R-symmetry group
$\SU(\cN)\times \U(1)$ \footnote{The $\U(1)$ factor being absent
for the case $\cN=8$.}, their position (lower or upper)
characterizing the left or right chirality of the gravitini, while
the index $I$, enumerating the spin-$1/2$ fields, is a short-hand
notation for the tensor character of the spin-$1/2$ fields.

The same matrices also appear in the order $g$ contribution to the
supersymmetry transformation laws of the fermions which, as it is
well known, is implied by the gauging procedure:
\begin{eqnarray}
\delta\lambda_I &=& \cdots +g\,N_I^A \,\epsilon_A \\
\delta\psi_{\mu A}&=& \cdots+{\rm i}\,g\, S_{AB}
\,\gamma_\mu\,\epsilon^B\,.
\end{eqnarray}
In an anti de Sitter background preserving all the $\cN$
supersymmetries we have:
\begin{eqnarray}
\delta\lambda_I &=& \cdots +\,g\,N_I^A \,\epsilon_A=0 \qquad \Rightarrow N_I^A|_{\mbox{\tiny SuSy AdS}}=0\\
\delta\psi_{\mu A}&=& \cdots +{\rm i}\,g\, S_{AB}
\,\gamma_\mu\,\epsilon^B\neq 0 \qquad \Rightarrow
S_{AB}|_{\mbox{\tiny SuSy AdS}}\neq 0 \,.
\end{eqnarray}
The precise relation, on the background, between the gravitino
mass
\begin{eqnarray}
m_{3/2}=2\,g\, \sqrt{S_{AB}\bar S^{AB}/\cN}
\end{eqnarray}
 and the scalar
potential  is then found from eq. \eq{ward}:
\begin{eqnarray}
V(\phi|_{\mbox{\tiny SuSy AdS}})= -3\,m^2_{3/2} =
\Lambda\label{condvac}
\end{eqnarray}
where $\Lambda$ is the cosmological constant.

Let us write down explicitly how the scalar potential specializes,
following from the above relations, for the $\cN=2$  and $\cN=1$
cases, and what are the conditions to have an anti de Sitter
vacuum with unbroken gauge symmetry and preserving all
supersymmetry. Note that the relations on the gauging of the
$\cN=1$ theory can also be obtained from the ones on
$\cN=2$-extended supergravity by a consistent truncation, as
discussed in \cite{Andrianopoli:2001zh}. For the $\cN=2$ theory,
in the absence of hypermultiplets, we find
\cite{Andrianopoli:1996cm}:
\begin{eqnarray}
V&=& -\frac 12 (\Im \cN^{-1})^{\Lambda\Sigma}\cP_\Lambda
\cP_\Sigma + (U^{\Lambda\Sigma} -3 \bar L^\Lambda L^\Sigma)
\cP^x_\Lambda \cP^x_\Sigma \label{n=2pot}
\end{eqnarray}
where $\cP_\Lambda$ is the prepotential for special geometry and
$\cP^x_\Lambda$ the constant quaternionic prepotential
corresponding to a Fayet--Iliopoulos term. The first term in eq.
\eq{n=2pot} is usually written $g_{i\bar\jmath}\,k^i_\Lambda\,
k^{\bar\jmath}_\Sigma \,\bar L^\Lambda L^\Sigma$, see for instance
\cite{Andrianopoli:1996cm}, where $k^i_\Lambda$ are the Killing
vectors of the special K\"ahler manifold and $L^\Lambda$ is the
upper part of the covariantly holomorphic symplectic  section.
Using $k^i_\Lambda =i\,
g^{i\bar\jmath}\partial_{\bar\jmath}\cP_\Lambda$, the
orthogonality relations $\cP_\Lambda L^\Lambda \equiv \cP_\Lambda
\bar L^\Lambda =0$ and the definition $U^{\Lambda\Sigma}\equiv
g^{i\bar\jmath}f^\Lambda_i \bar f^\Sigma_{\bar\jmath}= -\frac 12
(\Im \cN^{-1})^{\Lambda\Sigma}-\bar L^\Lambda L^\Sigma$, where
$f^\Lambda_i=\mathcal{D}_i\,L^\Lambda$ and $\cN_{\Lambda\Sigma}$
is the kinetic matrix of the vector fields, one easily retrieves
the expression in (\ref{n=2pot}) from the general one in
\cite{Andrianopoli:1996cm}.
 The
condition for an anti de Sitter supersymmetric background with
unbroken gauge group is
\begin{eqnarray}
\cP_\Lambda|_{vac}=0\,,\quad \left(U^{\Lambda\Sigma}\cP^x_\Lambda \cP^x_\Sigma\right)_{vac} =0\nonumber\\
\cP^x_\Lambda|_{vac} \neq 0
\end{eqnarray}

For the $\cN=1$ case, instead, the scalar potential has the general form \cite{Cremmer:1982en}:
\begin{eqnarray}
V&=& e^{\mathcal{K}} \left[\mathcal{D}_i W
\mathcal{D}_{\bar\jmath} \bar W \,g^{i{\bar\jmath}} -3|W|^2
\right] + \frac 12 (\Im f^{-1})^{AB} D_A D_B
\end{eqnarray}
where $f_{AB}$ denotes the holomorphic vector kinetic matrix,
$W(\phi)$ is the superpotential appearing in the fermion shifts of
the chiral multiplet fermions and $D_A$ is the D-term appearing in
the fermion shifts of the gaugini in the presence of gauged
isometries in the chiral multiplet sector. In this case, the
condition for an anti de Sitter vacuum preserving all
supersymmetries and gauge symmetry is
\begin {eqnarray}
D_A|_{vac}=0\,,\quad \mathcal{D}_i W|_{vac}=0 \nonumber\\
W|_{vac} \neq 0
\end{eqnarray}
The cosmological constant is then, in this case
\begin{eqnarray}
\Lambda = V|_{vac}=-3\,\left(e^{\mathcal{K}} \,|W|^2 \right)_{vac} \end{eqnarray}
and the gravitino mass is
\begin{eqnarray}
m_{3/2}= \left(|W|\,e^{\mathcal{K}/2} \right)_{vac}
\end{eqnarray}


\section{Dual  $\mathcal{N}=6$ and $\mathcal{N}=2$ gauged theories}\label{n6n2}
It is known that ungauged $\cN=6$ supergravity can be obtained from
ungauged $\cN=8$ supergravity by truncating out two gravitini
multiplets. At a group theoretical level this corresponds to
decomposing the relevant fermionic $\SU(8)$  representations with
respect to $\SU(6)\times \SU(2)\times \U(1)$, under which the ${\bf
8}$ branches as
\begin{eqnarray}
{\bf 8}&\rightarrow & {\bf (6,1)}_{+\frac{1}{2}}+{\bf
(1,2)}_{-\frac{3}{2}}\,,\label{8}
\end{eqnarray}
and keeping only the singlets under $\SU(2)$. In the following we
shall use the indices $i,j,\dots=1,\dots, 8$ to label the ${\bf
8}$ representation, which split into  indices $\alpha,
\beta,\dots=1,2$ labelling the ${\bf (1,2)}$ and
$A,B,\dots=1,\dots, 6$ labelling the ${\bf (6,1)}$\footnote{ Here
and in the following we reserve the indices $A,B,\dots$ only to
label the fundamental representation of the $\U(6)$ R-symmetry
group, while in the previous section they were associated with the
fundamental representation of the R-symmetry group of a generic
$\cN$--extended supergravity.}.
 Equation (\ref{8}) implies that the
$\mathcal{N}=8$ gravitini $\psi_\mu^i$ decompose under $
\SU(6)\times\SU(2)\times \U(1)\subset\SU(8)$, as $\psi_{i\,\mu}
\to(\psi_{A\,\mu}, \psi_{\alpha\,\mu})$, while  the spin $1/2$
fields $\chi_{ijk}$ into
 $(\chi_{ABC},
\chi_{AB\alpha},\chi_{A\alpha\beta}\equiv \chi_A
\epsilon_{\alpha\beta})$, according to the following branching
\begin{eqnarray}
{\bf 56}&\rightarrow&{\bf (20,1)}_{+\frac{3}{2}}+{\bf
(6,1)}_{-\frac{5}{2}}+{\bf (15,2)}_{-\frac{1}{2}}\,.\label{56}
\end{eqnarray}
The ${\bf 28}$ $\SU(8)$ representation of the $\cN=8$ central
charges $Z_{ij}$ branches in the following way
\begin{eqnarray}
{\bf 28}&\rightarrow & {\bf (15,1)}_{+1}+{\bf (1,1)}_{-3}+{\bf
(6,2)}_{-1}\,,\label{28}
\end{eqnarray}
where ${\bf (15,1)}_{+1}+{\bf (1,1)}_{-3}$, to be labelled by the
index $\underline{\Lambda}$, represent the $\cN=6$ central charges
$Z_{AB}$ and the singlet $Z_{\alpha\beta}=Z\,\epsilon_{\alpha\beta}$, while
 the remaining charges in the ${\bf (6,2)}_{-1}$ are truncated.

The corresponding branching of the $\SU(8)$ representation ${\bf
70}$ pertaining to the scalar fields $\phi^{ijkl}$, spanning
$\sM_{(\cN=8)}=\E_{7(7)}/\SU(8)$, reads:
\begin{eqnarray}
{\bf 70}&\rightarrow & {\bf (15,1)}_{-2}+{\bf
(\overline{15},1)}_{+2}+{\bf (20,2)}_{0}\,.
\end{eqnarray}
The truncation to the $\SU(2)$ singlets yields the $30$ scalar
fields of the $\cN=6$ theory which span the coset manifold
\begin{eqnarray}
\sM_{(\cN=6)} &=&\frac{\SO^*(12)}{\U(6)}\,,\label{scalma}
\end{eqnarray}
which is a submanifold of $\sM_{(\cN=8)}$ . The global on-shell
symmetry group of the theory is $\SO^*(12)$ which acts as a
generalized electric-magnetic duality. The $32$ electric-magnetic
charges are indeed obtained by branching the $\E_{7(7)}$
representation ${\bf 56}$ of the corresponding $\cN=8$ charges
with respect to the maximal subgroup $\SO^*(12)\times \SU(2)$ of
$\E_{7(7)}$ and keeping only the singlets:
\begin{eqnarray}
{\bf 56}&\rightarrow & {\bf (12,2)}+{\bf (32,1)}\,.
\end{eqnarray}
If in the $\cN=8$ theory we truncate the multiplets of the six
gravitini fields $\psi_{A\mu}$ instead, we would obtain an $\cN=2$
theory with the same bosonic sector as the
 $\cN=6$ model but a different
 fermionic field content. The $\cN=8$ central charges give now rise to the $\cN=2$ central charge
 $Z$ and 15 matter charges $Z_{AB}$.
 This theory therefore describes $\cN=2$ supergravity
 coupled to 15 vector multiplets and no hypermultiplets. The scalar fields in the vector multiplets span the special K\"ahler manifold (\ref{scalma}).
The spin $3/2$ fields $\psi_{\alpha\mu}$ belong to the ${\bf
(1,2)}_{-\frac{3}{2}}$ representation in (\ref{8}) while the spin
$1/2$ fields $\chi_{AB\alpha}$ are defined by the ${\bf
(15,2)}_{-\frac{1}{2}}$ representation in the branching
(\ref{56}).   This peculiarity of the $\cN=2$ and $\cN=6$
truncations just discussed, to share the same bosonic content
although differing in the fermionic sector, was exploited in the
study of extremal black holes, where one finds a class of common
extremal solutions, which, however, have different supersymmetry
properties in the two theories: The BPS solution of the $\cN=6$
theory is non-BPS in the $\cN=2$ one and vice versa
\cite{Andrianopoli:1996ve,Ferrara:2006yb,Bellucci:2006xz,Ferrara:2006xx,Andrianopoli:2006ub}.

To
summarize the $\cN=8\rightarrow \cN=6,\,\cN=2$ truncations
discussed above, let us  denote by $\Phi_{(in)}$ the (bosonic and
fermionic) fields surviving the truncation and by $\Phi_{(out)}$
those fields which are truncated away. For the two truncations
these fields read:
\begin{eqnarray}
\mbox{$\cN=6$}&:&\begin{cases}\Phi_{(in)}=\{
\phi^{AB\alpha\beta}=\phi^{AB}\epsilon^{\alpha\beta},
A^{\alpha\beta}_\mu,\,A^{AB}_\mu,\,\psi^A_\mu,\,\chi^{ABC},\,\chi^{A},\,c.c\}\cr
\Phi_{(out)}=\{\phi^{ABC\beta},\,
A^{A\alpha}_\mu,\,\psi^\alpha_\mu,\,\chi^{AB\alpha},\,c.c.\}\end{cases}\,,\\
\mbox{$\cN=2$}&:&\begin{cases}\Phi_{(in)}=\{
\phi^{AB\alpha\beta}=\phi^{AB}\epsilon^{\alpha\beta},
A^{\alpha\beta}_\mu,\,A^{AB}_\mu,\,\psi^\alpha_\mu,\,\chi^{AB\alpha},\,c.c\}\cr
\Phi_{(out)}=\{\phi^{ABC\beta},\,
A^{A\alpha}_\mu,\,\psi^A_\mu,\,\chi^{ABC},\,\chi^{A},\,c.c.\}\end{cases}
\end{eqnarray}
\par Let us now consider the gauging of these $\cN=6$
and $\cN=2$ theories. As we shall show such gauged theories can
all be constructed as a truncation of the $\cN=8$ theory with a
suitable gauging. The most general $\cN=8$ gauged supergravity can
be written in a manifestly $\SU(8)$ invariant form \cite{de
Wit:1982ig}, in which the fermion shifts, which define the fermion
mass terms and the scalar potential, consist in a symmetric tensor
$S_{ij}=S_{ji}$ and a tensor $N^i{}_{jkl}$ in the ${\bf 36}$ and
${\bf 420}$ of $\SU(8)$ respectively\footnote{It is useful here to
define the correspondence between our notation and the one used in
\cite{dwst1,dwst2,dwst3}, to be distinguished by a prime from the
quantities denoted here with the same symbol:
$\gamma^\mu=i\,\gamma^{\prime\mu}$,
$\psi^i_\mu=\frac{1}{\sqrt{2}}\,\psi^{\prime\, i}_\mu$,
$\epsilon^i=\sqrt{2}\,\epsilon^{\prime\, i}$,
$S^{ij}=-\frac{1}{\sqrt{2}}\,A_1^{ij}$,
$N_\ell{}^{ijk}=-\sqrt{2}\,A_2{}_\ell{}^{ijk}$}. In terms of these
quantities, the supersymmetry variations of the (chiral components
of the) fermion fields read:
\begin{eqnarray}
\delta
\psi_\mu^i&=&\cdots +i\,g\,S^{ij}\,\gamma_\mu\,\epsilon_j\,,\\
\delta \chi^{ijk}&=&\cdots+g\,N_{l}{}^{ijk}\,\epsilon^l\,.
\end{eqnarray}
 According to the
general form (\ref{ward}) of the Ward identity, the $\cN=8$ scalar
potential reads:
\begin{eqnarray}
V^{(\cN=8)}(\phi)&=&g^2\,\left(\frac{1}{48}\,N_{i}{}^{jkl}\,N^i{}_{jkl}-\frac{3}{2}\,S^{ij}\,S_{ij}\right)\,.\end{eqnarray}
  As far as the supersymmetry transformation rules
are concerned, for the order $g$ sector involving the fermion
shifts we find the decomposition:
\begin{eqnarray}
\delta \psi_\mu^A&=&\cdots +
i\,g\,\left(S^{AB}\,\gamma_\mu\,\epsilon_B+
S^{A\beta}\,\gamma_\mu\,\epsilon_\beta\right)
\,,\label{psia}\\
\delta \psi_\mu^\alpha&=&\cdots + i\,g\,\left(S^{\alpha
B}\,\gamma_\mu\,\epsilon_B +S^{\alpha \beta}\,
\gamma_\mu\,\epsilon_\beta\right)\,,\label{psial}
\end{eqnarray}
and
\begin{eqnarray}
\delta \chi^{ABC}&=&\cdots+g\,\left(N_{D}{}^{ABC}\,\epsilon^D+N_{\beta}{}^{ABC}\,\epsilon^\beta\right)\,,\label{chiabc}\\
\delta \chi^{AB\alpha}&=&\cdots+g\,\left(N_{D}{}^{
AB\alpha}\,\epsilon^D+N_{\beta}{}^{
AB\alpha}\,\epsilon^\beta\right)\,,\label{chiabal}
\\
 \delta\chi^{A}&=&\cdots+\,g\,\left(N_{B}{}^{A}\,\epsilon^B+N_{\beta}{}^{A}\,\epsilon^\beta\right)\,.\label{chia}
\end{eqnarray}
The above fermion shifts correspond respectively to the
branchings:
\begin{eqnarray}
{\bf 36}&\rightarrow &  {\bf (21,1)}_{+1}+{\bf (6,2)}_{-1}+{\bf
(1,3)}_{-3}\label{36}\\
{\bf 420}&\rightarrow &  {\bf (105,1)}_{+1} +{\bf
(20,2)}_{+3}+{\bf (84,2)}_{-1}+{\bf (15,1)}_{+1}+{\bf (15,3)}_{+1}
+{\bf (35,1)}_{-3} +{\bf (6,2)}_{-1}\,.\nonumber\\\label{420}
\end{eqnarray}
In order to have a consistent truncation, the solutions of the
equations of motion of the reduced theory must also  be solution
in the parent theory, namely
\begin{eqnarray}
\frac{\delta \mathcal{L}}{\delta\Phi_{(out)}}&\approx &0\,,
\end{eqnarray}
where $\approx 0$ have to be intended in a weak sense, namely at
$\Phi_{(out)}\equiv 0$. This fact in particular implies that all
terms in the Lagrangian bilinear in the fermions and containing
one retained and one truncated fermion, must disappear in the
reduction, otherwise the corresponding field equations obtained by
varying the Lagrangian with respect to the truncated fermions,
would not be (weakly) satisfied. Let us consider the following
order $g$ fermion bilinears in the gauged $\cN=8$ Lagrangian,
which can be derived from the general expression for the fermion
mass-like terms (\ref{psim}) and (\ref{chim})\footnote{Here we
restrict to the $\psi\psi$ and $\psi\chi$ terms in the Lagrangian.
We refer the reader to Appendix \ref{masst} for the explicit form
of the $\chi\chi$ mass-like terms.}:\begin{eqnarray} &&g\,\left(
4\,S_{A\alpha} \,\bar \psi^A_\mu \,\gamma^{\mu\nu}\psi^\alpha_\nu+
\frac{1}{6}\,N^\alpha{}_{BCD}\,\bar
\chi^{BCD}\,\gamma^{\mu}\psi_{\alpha\,\mu
}+\right.\nonumber\\&&\qquad\qquad\left.+\frac{1}{2}\,N^A{}_{BC\alpha}
\,\bar \chi^{BC\alpha}\,\gamma^{\mu}\psi_{A\,\mu}+
N^\alpha{}_{ A} \,\bar \chi^{A}\,\gamma^{\mu}\psi_{\alpha\,\mu
}\right) \,+ h.c.\,,\nonumber\\&&\label{bilinlin}
\end{eqnarray}

%
%
Since the terms in  eq. (\ref{bilinlin}) are linear in the
truncated fermions, we conclude that consistency of the two
truncations requires  the components of the $\cN=8$ fermion shifts
which transform as doublets of $\SU(2)$, namely $S_{\alpha B}$ in
the ${\bf (6,2)}_{-1}$, $N^{\beta}{}_{ABC}$ in the ${\bf
(20,2)}_{+3}$, $N^{D}{}_{ AB\alpha}$ in the ${\bf (84,2)}_{-1}$
and $N^{\beta}{}_{A}$ in the ${\bf (6,2)}_{-1}$, to be weakly
zero. Therefore in order for the truncation of gauged $\cN=8$ to
$\cN=6$ or $\cN=2$ to be consistent, the gauging must be such
that, when restricted to the common scalar sector of the two
truncations, the components of the fermion shifts transforming as
doublets under $\SU(2)$ must vanish. From now on we shall assume
this to be the case.  The implications of this condition  on the
possible gauge groups will be discussed in the next subsection.
The resulting $\mathcal{N}=6$ and $\mathcal{N}=2$ theories then
involve the transformation rules:
\begin{eqnarray}
\delta
\psi_\mu^A&=&\cdots +i\,g\,S^{AB}\,\gamma_\mu\,\epsilon_B\,,\label{61}\\
\delta \chi^{ABC}&=&\cdots+g\,N_{D}{}^{ABC}\,\epsilon^D\,,\label{62}\\
 \delta\chi^{A}&=&\dots+g\,N_{B}{}^{A}\,\epsilon^B\,,\label{63}
\end{eqnarray}
 for the $\cN=6$ theory, while for the $\cN=2$ theory we have:
\begin{eqnarray}
\delta
\psi_\mu^\alpha&=&\cdots +i\,g\,S^{\alpha\beta}\,\gamma_\mu\,\epsilon_\beta\,,\label{21}\\
\delta \chi^{AB\alpha}&=&\cdots+g\,N_{\beta}{}^{\alpha
AB}\,\epsilon^\beta\,.\label{22}
\end{eqnarray}
While $S_{AB}$, $S_{\alpha\beta}$, $N_{B}{}^{A}\equiv
N_{B}{}^{A\alpha\beta}\,\epsilon_{\alpha\beta}/2$ are irreducible
$\SU(6)\times \SU(2)\times \U(1)$-tensors in the ${\bf
(21,1)}_{+1}$, ${\bf (1,3)}_{-3}$ and ${\bf (35,1)}_{+3}$
respectively, the shift tensors
$N_{D}{}^{ABC},\,N_{\beta}{}^{\alpha AB}$ transform in reducible
representations and can therefore be written as follows:
\begin{eqnarray}
N_{D}{}^{ABC}&=&\oN_{D}{}^{ABC}+\frac{3}{4}\,\delta^{[A}_D\,N^{BC]}\,,\nonumber\\
N_{\beta}{}^{\alpha AB}&=&\oN_{\beta}{}^{\alpha
AB}-\frac{1}{2}\,\delta^{\alpha}_\beta\,N^{AB}\,,\label{decompN}
\end{eqnarray}
where the irreducible tensors
$\oN_{D}{}^{ABC},\,N^{AB},\,\oN_{\beta}{}^{\alpha AB}$ transform
in the ${\bf (\overline{105},1)}_{-1}$, ${\bf
(\overline{15},1)}_{-1}$ and  ${\bf (\overline{15},3)}_{-1}$
representations respectively (this implies in particular the
properties
$\oN_{D}{}^{ABC}=\oN_{D}{}^{[ABC]},\,\oN_{C}{}^{ABC}=0$).\par
 Let us note, however, that the shifts involved in the
transformation of projected out gravitini and dilatini with
respect to projected out supersymmetry parameters are in general
different from zero. In the truncation to $\mathcal{N}=6$ they are
$S^{\alpha \beta}$ and $N_{\beta}{}^{\alpha AB}$, while in the
truncation to $\mathcal{N}=2$ they are $S^{AB}$,  $N_{D}{}^{ABC}$
 and
$N_{B}{}^{A\alpha\beta}=N_{B}{}^{A}\,\epsilon^{\alpha\beta}$. Some
of them still enter the Lagrangian, as it is the case for the
$\oN_{D}{}^{ABC}$ component of $N_{D}{}^{ABC}$, which enters the
fermion mass term in the $\cN=2$ theory, or the $N^{AB}$ component
of $N_{\beta}{}^{\alpha AB}$ which enters both in the shift tensor
$N_{D}{}^{ABC}$ and in the spin-$1/2$ mass terms of the $\cN=6$
truncation (see Appendix \ref{masst}). These shifts moreover play
a role in rewriting the $\mathcal{N}=8$ scalar potential in terms
of the only fermion shifts pertaining to the two truncations. The
simplest way to achieve this is perhaps to restrict the
$\mathcal{N}=8$ Ward identity:
\begin{eqnarray}
 \delta^i_j V^{(\cN=8)}= g^2\,\left(-12\,S^{ik}\,S_{
jk} + \frac{1}{6}\,N_{j}{}^{ k\ell m} N^{ i}{}_{k\ell
m}\right)\,,\label{ward8}
\end{eqnarray}
to the $\cN=6$ and $\cN=2$ indices and to the common scalar
content of the two truncations:
\begin{eqnarray}
 \delta^A_B\, V^{(\cN=8)}&\approx& g^2\,\left(-12\,S^{AC}\,S_{
BC} + \frac{1}{6}\,N_{B}{}^{ CDE} N^{ A}{}_{CDE}+N_{B}{}^{ C} N^{
A}{}_{C}\right)\,,\label{ward6}\\
 \delta^\alpha_\beta\, V^{(\cN=8)}&\approx& g^2\,\left(-12\,S^{\alpha
 \gamma}\,S_{\beta\gamma} + \frac{1}{2}\,N_{\beta}{}^{ \gamma AB} N^{
\alpha}{}_{\gamma AB}\right)\,,\label{ward2}
\end{eqnarray}
where, as usual $\approx$ denotes the restriction to
$\Phi_{(in)}$. By tracing the above identities we obtain the
scalar potential written in terms of $\cN=6$ and $\cN=2$
quantities respectively:
\begin{eqnarray}
  V^{(\cN=8)}&\approx& V^{(\cN=6)}= g^2\,\left(-2\,S^{AB}\,S_{
AB} + \frac{1}{36}\,N_{A}{}^{ BCD} N^{
A}{}_{BCD}+\frac{1}{6}\,N_{A}{}^{ B} N^{
A}{}_{B}\right)\,,\label{pot6}\\
  V^{(\cN=8)}&\approx& V^{(\cN=2)}= g^2\,\left(-6\,S^{\alpha
 \beta}\,S_{\alpha
 \beta} + \frac{1}{4}\,N_{\alpha}{}^{ \beta AB} N^{
\alpha}{}_{\beta AB}\right)\,.\label{pot2}
\end{eqnarray}
Note that the two expressions (\ref{pot6}),(\ref{pot2}) are
alternative descriptions of a same functional, which is the
restricted $\cN=8$ potential. We conclude that the $\cN=8$ Ward
identity implies a non trivial relation between the $\cN=6$ and
$\cN=2$ fermion shifts, which is crucial in order to rewrite the
same restricted $\cN=8$ potential in terms of the quantities
pertaining to the two truncations.

\subsection{The gaugings of the $\cN=6$ and $\cN=2$ truncations}
Having discussed  the general form of the $\cN=6$ and $\cN=2$
truncations of the  (gauged) $\cN=8$ theory, let us show that
these describe respectively the most general gauged $\cN=6$ theory
and the most general $\cN=2$ gauged supergravity, based on the
scalar manifold (\ref{scalma}). In other words we consider here
the problem of characterizing the most general local symmetries
which these models may exhibit.
 To this end it is useful to describe their
gauging by using the \emph{embedding tensor} formalism
\cite{dwst1,dwst2,dwst3} (for recent reviews on the embedding tensor
formalism and its application to flux compactifications see
\cite{dwstrev}). Let us briefly recall the main facts about this
technique and consider the gauging of an extended supergravity with
$n_v$ vector fields $A^\Lambda_\mu$, $\Lambda=1,\dots, n_v$, and a
 scalar manifold of the form $G/H$, where $G$ represents the
on-shell (classical) global symmetry group and $H$ its maximal
compact subgroup. The gauging procedure consists in promoting a
suitable subgroup $\mathcal{G}$ of the global symmetry group of the
Lagrangian to local symmetry, gauged by (a subset of) the electric
potentials of the theory. The formalism introduced in
\cite{dwst2,dwst3} allows to freely choose the candidate gauge group
inside the full on-shell global symmetry group $G$ of the ungauged
theory by allowing the minimal couplings to involve not just the
electric fields but also the magnetic ones $A_{\Lambda\,\mu}$ in a
symplectic covariant fashion\footnote{ Consistency of the
construction also requires the addition of antisymmetric tensor
fields in the adjoint representation of $G$. Additional gauge
symmetries guarantee that the introduction of these extra fields
does not add new degrees of freedom to the theory.} . In this way
the analysis of all possible gaugings is no longer constrained by
the choice of the original ungauged Lagrangian and can refer to the
full non-perturbative symmetries of the ungauged theory. Let us use
the index $M$ to label the symplectic representation ${\bf R}$ of
$G$ in which the electric and magnetic charges transform, so that a
generic symplectic vector reads $V^M=(V^\Lambda,\,V_\Lambda)$. We
shall also denote by $\Omega_{MN}$ the symplectic invariant matrix.
Finally let the index $n$ label the adjoint representation of $G$.
The choice of the gauge algebra inside the Lie algebra of $G$, to be
gauged by a subset of the electric and magnetic potentials, can be
parametrized by a $G$-covariant embedding tensor $\theta_M{}^n$,
which expresses the gauge generators $X_M$ as a linear combination
of the generators $t_n$ of $G$: $X_M=\theta_M{}^n\,t_n$. By
definition $\theta_M{}^n$ naturally belongs to the product ${\bf
R}\times {\bf Adj}(G)$. The deformations of the original ungauged
Lagrangian which yield the gauged one with the same amount of
supersymmetries, can be written in terms of the embedding tensor in
a $G$-invariant way. Consequently  the gauged equations of motion
and Bianchi identities formally exhibit the same global symmetries
as the ungauged ones provided $\theta_M{}^n$ is transformed under
$G$ as well. This action of $G$ extended to $\theta_M{}^n$ can be
interpreted as a mapping between  different gauged supergravities.
The electric-magnetic duality action of the generators $t_n$ of $G$
is represented by symplectic matrices $(t_n)_M{}^P$, which are meant
to act on the vectors of electric and magnetic charges. We can then
define the $G$-tensor $X_{MN}{}^P=\theta_M{}^n\,(t_n)_N{}^P$, in the
same representation as $\theta_M{}^n$. For theories with $\cN\le 2$
not all generators of $G$ are associated with an electric-magnetic
duality action (as it is the case for the quaternionic isometries in
$\cN=2$ theories). These symmetries have $(t_{n})_M{}^N=0$ and thus
do not contribute to $X_{MN}{}^P$. Consistency of the construction
of a gauged extended supergravity requires $\theta_M{}^n$ to satisfy
some $G$-covariant constraints consisting of a linear condition on
$X_{MN}{}^P$:
\begin{eqnarray}
X_{(MN}{}^L\,\Omega_{P)L}&=&0\,,\label{1st}
\end{eqnarray}
and the following quadratic conditions
\begin{eqnarray}
\theta_M{}^m\,\theta_N{}^n\,f_{mn}{}^p+X_{MN}{}^P\,\theta_{P}{}^p&=&0\,,\label{2nd1}\\
\theta_M{}^m\theta_N{}^n\,\Omega^{MN}&=&0\label{2nd2}\,,
\end{eqnarray}
where $f_{mn}{}^p$ are the structure constants of $G$:
$[t_n,\,t_m]=f_{mn}{}^p\,t_p$. Equation \eq{2nd1} expresses the
requirement that $\theta_M{}^n$ be a gauge invariant quantity and
implies the closure of the gauge algebra $\frak{g}$ inside the Lie
algebra of $G$: $[X_M,\,X_N]=-X_{MN}{}^P\,X_P$. Equation
(\ref{2nd2}) guarantees mutual locality between the electric and
magnetic components of $\theta_M{}^n$. In supergravities with $\cN>
2$ all $t_n$ have non trivial electric-magnetic duality action and
it can be shown that (\ref{1st}) and (\ref{2nd1}) imply
(\ref{2nd2}). The quadratic conditions (\ref{2nd1}), (\ref{2nd2}) on
the structure constants of the gauge algebra imply the Ward identity
(\ref{ward}) which is crucial for the supersymmetry of the gauged
Lagrangian.
\par
 Note that the
constraints  (\ref{1st}), (\ref{2nd1})  and (\ref{2nd2}) are
manifestly $G$-covariant. The linear one in particular amounts to a
condition on $G$-representation of the embedding tensor in the
decomposition of ${\bf R}\times {\bf Adj}(G)$. For instance in the
maximal theory $G=\E_{7(7)}$, $H=\SU(8)$,  ${\bf R}={\bf 56}$, ${\bf
Adj}(G)={\bf 133}$ and (\ref{1st}) implies that $\theta_M{}^n$
belong to the ${\bf 912}$ representation in the decomposition of
${\bf 56}\times{\bf 133}$.\par
 As far as the $\cN=6$ and $\cN=2$ truncations are concerned,
 in both cases the global symmetry group $G$
can be identified with the maximal subgroup $\SO^*(12)\times
\SU(2)$ of $\E_{7(7)}$, with the only difference that in the
former theory the $\SU(2)$ has a trivial action since all fields
are singlets with respect to it, while this is not the case for
the latter model. In the $\cN=2$ truncation the $\SU(2)$ factor is
a global symmetry group whose generators $t_x$, $x=1,2,3$, have a
trivial electric-magnetic duality action: $(t_x)_M{}^N=0$. As we
shall see the gauging of this $\SU(2)$ group amounts to
introducing a Fayet-Iliopuolos term. \par In both the $\cN=6$ and
$\cN=2$ theories, ${\bf R}={\bf (32,1)}$, ${\bf Adj}(G)={\bf
(66,1)}+{\bf (1,3)}$ and the decomposition of ${\bf R}\times {\bf
Adj}(G)$ reads
\begin{eqnarray}
{\bf (32,1)}\times[{\bf (66,1)}+{\bf (1,3)}]&\rightarrow & {\bf
(32,1)}+{\bf (1728,1)}+{\bf (352,1)}+{\bf (32,3)}\,.
\end{eqnarray}
The constraint (\ref{1st}) implies that the representations in the
above decomposition which are in common with the three times
symmetric product of the ${\bf (32,1)}$ should vanish. Since
\begin{eqnarray}
[{\bf (32,1)}\times {\bf (32,1)}\times {\bf
(32,1)}]_{sym.}&\rightarrow & {\bf (32,1)}+{\bf (4224,1)}+{\bf
(1728,1)}\,,
\end{eqnarray}
we conclude that in both theories the most general gaugings are
defined by an embedding tensor in the following representations:
\begin{eqnarray}
\theta_M{}^n&\in& {\bf (352,1)}+{\bf (32,3)}\,.\label{n26emb}
\end{eqnarray}
The gaugings parametrized by an embedding tensor $\theta_M{}^x$ in
the ${\bf (32,3)}$ representation involve the $\SU(2)$ generators
and therefore have no effect in the $\cN=6$ theory. In the $\cN=2$
theory instead they correspond to introducing an electric-magnetic
F-I term, corresponding to constant electric and magnetic momentum
maps $\mathcal{P}^x_M=(\mathcal{P}^x_\Lambda,
\mathcal{P}^{x\Lambda})\equiv \theta_M{}^x$. Condition (\ref{2nd1})
in this case expresses the \emph{equivariance} of the (constant)
momentum maps:
\begin{eqnarray}
\mathcal{P}_M{}^x\,\mathcal{P}_N{}^y\,\epsilon_{xy}{}^z+X_{MN}{}^P\mathcal{P}_{P}{}^z&=&0\,,\label{equiv}\,.
\end{eqnarray}
 Note that the
representations (\ref{n26emb}) occur in the branching of the ${\bf
912}$ of $\E_{7(7)}$ with respect to $\SO^*(12)\times \SU(2)$
\begin{eqnarray}
{\bf 912}&\rightarrow &{\bf (12,2)}+{\bf (220,2)}+{\bf
(352,1)}+{\bf (32,3)}\,,\label{dec912}
\end{eqnarray}
and are the only non-doublet representations. From this we
conclude that \emph{the most general $\cN=6$ gauged supergravity
can be obtained from the gauged $\cN=8$ supergravity by truncating
the fields and the embedding tensor to the singlet representations
with respect to $\SU(2)$}. Let us illustrate the implications of
the above discussion on the fermion shifts and scalar potential of
the gauged $\cN=6$ supergravity.
\par
In a generic gauged extended supergravity, the fermion  shifts,
which  belong to representations of $H$,  are linear in the
embedding tensor. In an extended supergravity based on a homogeneous
symmetric scalar manifold, they are in fact expressed in terms of
the so called T-tensor (originally introduced in \cite{de
Wit:1982ig} for the maximal supergravity), which is an
$H$--covariant quantity, obtained by ``boosting'' $\theta_M{}^n$ by
means of the scalar-dependent coset representative
$\mathcal{V}(\Phi)$:
\begin{eqnarray}
T(\Phi,\theta)_{\underline{M}}{}^{\underline{n}}&= &
(\mathcal{V}^{-1}\star
\theta)_{\underline{M}}{}^{\underline{n}}\equiv
\mathcal{V}^{-1}{}_{\underline{M}}{}^M
\,\mathcal{V}_n{}^{\underline{n}}\,\theta_M{}^n\,,\label{ttensor0}
\end{eqnarray}
where $\mathcal{V}_M{}^{\underline{M}}$ and
$\mathcal{V}_n{}^{\underline{n}}$ are the matrix representations of
the coset representative  in the ${\bf R}$ and ${\bf Adj}(G)$
representations of $G$, while the underlined indices are acted on by
$H$ transformations. If the scalar fields $\Phi$ and $\theta_M{}^n$
are simultaneously transformed  by means of a $G$ transformation
${\bf g}$, $T(\Phi,\theta)$ transforms  under a corresponding
$H$--compensating transformation depending on $\Phi$ and ${\bf g}$.
In this sense $T(\Phi,\theta)$ is an $H$ covariant quantity, and
thus can be decomposed into irreducible $H$-- representations. These
irreducible components comprise the fermion shift tensors. However
$T(\Phi,\theta)$ can also be viewed as a $G$-tensor, since it is
obtained by acting on the $G$-tensor $\theta$ by means of a
$G$-transformation $\mathcal{V}(\Phi)$. This implies that
$T(\Phi,\theta)$ satisfies the same linear and quadratic constraints
as $\theta$ and thus, in particular, that it should belong to the
same $G$-representation as $\theta$. The quadratic constraints on
$T(\Phi,\theta)$, on the other hand, imply the Ward identity for the
fermion  shifts.
 Therefore the
$H$- representations defining the fermion shift tensors should
appear in the branching of the embedding tensor (or T-tensor)
$G$--representation with respect to $H$. For instance, in the
$\cN=8$ theory, the branching of the ${\bf 912}$ with respect to
$\SU(8)$ yields the $\SU(8)$--representations pertaining to $S^{ij}$
and $N_l{}^{ijk}$:
\begin{eqnarray}
{\bf 912}&\rightarrow &{\bf 36}+{\bf 420}+{\bf \overline{36}}+{\bf
\overline{420}}\,.
\end{eqnarray}

Similarly, for the $\cN=2$ and $\cN=6$ theories,  branching the
common embedding tensor representation (\ref{n26emb}) with respect
to the compact symmetry group $\SU(6)\times \SU(2)\times \U(1)$ we
find
\begin{eqnarray}
{\bf (352,1)}+{\bf (32,3)}&\rightarrow & {\bf (35,1)}_{+3}+{\bf
(21+15+105,1)}_{+1}+{\bf
(\overline{21}+\overline{15}+\overline{105},1)}_{-1}+\nonumber\\&&+{\bf
(\overline{35},1)}_{-3}+{\bf (1,3)}_{+3}+{\bf (15,3)}_{+1}+{\bf
(1,3)}_{-3}+{\bf (\overline{15},3)}_{-1}\,.\label{br352}
\end{eqnarray}
The correspondence of the above representations with the fermion
shifts introduced in \eq{61} - \eq{22}  is:
\begin{eqnarray}
\mbox{{\bf $\cN=6$}:}&&\qquad {\bf (35,1)}_{+3}\equiv
N_{B}{}^{A}\,\,,\,\,\,{\bf
(\overline{21}+\overline{105}+\overline{15},1)}_{-1}\equiv
(S^{AB},N_{D}{}^{ABC}\,)\,,\\
\mbox{{\bf $\cN=2$}:}&&\qquad {\bf (1,3)}_{+3}\equiv
S^{\alpha\beta}\,\,,\,\,\,{\bf (\overline{15},1)}_{-1}+{\bf
(\overline{15},3)}_{-1}\equiv N_{\beta}{}^{\alpha AB}\,.
\end{eqnarray}


\subsection{$\cN=6$ with $\SO(6)\times\SO(2) $ gauge group}
We shall now discuss $\cN=6$ gaugings in some detail and focus on
the theory with $\SO(6)\times\SO(2)$ local symmetry (the $\SO(2)$
factor, being contained in the $\SU(2)$ global symmetry, has a
trivial action on the $\cN=6$ fields). We start defining the
relation between the fermion shifts and the embedding tensor. Let
$\mathcal{V}_M{}^{\underline{M}}$ denote the coset representative
of the scalar manifold (\ref{scalma}):
\begin{eqnarray}
\mathcal{V}_M{}^{\underline{M}}&=&\left(\begin{matrix}\bar{h}_\Lambda{}^{\underline{\Lambda}}&h_{\Lambda\underline{\Lambda}}\cr
\bar{f}^{\Lambda\underline{\Lambda}}&f^\Lambda{}_{\underline{\Lambda}}&\end{matrix}\right)\,,\label{section}
\end{eqnarray}
where the underlined indices label the $\U(6)$ representations in
which the self dual and anti-self dual field strengths transform,
and the blocks ${\bf f}\equiv
(f^\Lambda{}_{\underline{\Lambda}}),\,\bar{{\bf f}}\equiv
(\bar{f}^{\Lambda\underline{\Lambda}}),\,{\bf h}\equiv
(h_{\Lambda\underline{\Lambda}}),\,\bar{{\bf h}}\equiv
(\bar{h}_\Lambda{}^{\underline{\Lambda}})$ satisfy the relations:
\begin{eqnarray}
({\bf f}\,{\bf f}^\dagger)^T&=&{\bf f}\,{\bf
f}^\dagger\,\,\,,\,\,\,\,({\bf h}\,{\bf h}^\dagger)^T={\bf h}\,{\bf
h}^\dagger\,\,\,,\,\,\,\,\, {\bf f}\,{\bf h}^\dagger-\bar{{\bf
f}}\,{\bf h}^T=i\,\bfone\,,\\{\bf f}^\dagger\,{\bf h}-{{\bf
h}}^\dagger\,{\bf f}&=&-i\,\bfone\,\,\,,\,\,\,\,{\bf f}^T\,{\bf
h}-{{\bf h}}^T\,{\bf f}={\bf 0}\,.
\end{eqnarray}
Using the above properties we can write the general expression of
$\mathcal{V}^{-1}$:
\begin{eqnarray}
\mathcal{V}^{-1}{}_{\underline{M}}{}^M&=&\left(\begin{matrix}-i\,f^\Lambda{}_{\underline{\Lambda}}&i\,h_{\Lambda\underline{\Lambda}}\cr
i\,\bar{f}^{\Lambda\underline{\Lambda}}&-i\,\bar{h}_\Lambda{}^{\underline{\Lambda}}\end{matrix}\right)\,.
\end{eqnarray}
The basic quantity in terms of which the fermion shifts are
expressed is the T--tensor, introduced in the previous section.
Since in the $\cN=6$ theory all the generators of  $G$ have a non
trivial duality action, the gauging is totally characterized by the
generalized structure constants $X_{MN}{}^P$. It is then convenient
here to use a slightly different definition of the T-tensor, with
respect to eq. (\ref{ttensor0}), and construct it by \emph{dressing}
$X_{MN}{}^P$ with the scalar fields by means of the coset
representative:
\begin{eqnarray}
T_{\underline{M},\underline{N}}{}^{\underline{P}}&=&[\mathcal{V}^{-1}\star
X]_{\underline{M},\underline{N}}{}^{\underline{P}}\equiv
\mathcal{V}^{-1}{}_{\underline{M}}{}^M\,\mathcal{V}^{-1}{}_{\underline{N}}{}^N\,\mathcal{V}_P{}^{\underline{P}}
\,X_{MN}{}^P\,.\label{TT}
\end{eqnarray}
To write the fermion shifts in terms of the above quantity, we can
use the corresponding $\cN=8$ relations and reduce them to the
$\cN=6$ theory. In the maximal gauged supergravity the following
relation holds:
\begin{eqnarray}
T_{ij,kl}{}^{pq}&=&-\frac{1}{2\sqrt{2}}\,\delta^{[p}_{[k}\,N^{q]}{}_{l]ij}-\sqrt{2}\,\delta^{[p}_{[k}\,S_{l][i}\,\delta^{q]}_{j]}\,.
\end{eqnarray}
 We then find:
\begin{eqnarray}
N^A{}_B&=&-2\,\sqrt{2}\, T_{\alpha\beta,\,BC}{}^{AC}\,,\,\,N_{AB}=
-\frac{8}{3}\,\sqrt{2}\,T_{C[A,B]E}{}^{CE}\,,\nonumber\\
N^A{}_{BCD}&=&-2\,\sqrt{2}\,T_{[CD,B]E}{}^{AE}-\frac{1}{4}\,
\delta^A_{[B}\,N_{CD]}
\,,\,\,S_{AB}=\frac{\sqrt{2}}{5}\,T_{C(A,B)E}{}^{CE}\,.
\end{eqnarray}
Let us now consider the gauging of $\mathcal{G}=\SO(6)$. Since the
embedding tensor, by construction,   defines the gauge structure
constants, it  is itself a gauge invariant quantity, as expressed by
eq. (\ref{2nd1}). This allows to define the embedding tensor
corresponding to a given gauge group $\mathcal{G}$ by considering
the singlets in the branching of the embedding tensor
$G$--representation with respect to $\mathcal{G}$. In particular the
embedding tensor corresponding to  $\mathcal{G}=\SO(6)$ must be
defined by a singlet in the branching of (\ref{n26emb}) with respect
to the $\SO(6)$ maximal subgroup of $\SU(6)$.  This singlet arises
only from the ${\bf 21}$ and $\overline{{\bf 21}}$ in the branching
(\ref{br352}): ${\bf 21}\rightarrow {\bf 20}+{\bf 1}$. The $\SO(6)$
generators are gauged by the electric potentials which transform its
adjoint representation, labelled by the antisymmetric couple $[IJ]$,
$I,J=1,\dots, 6$. The index $\Lambda$ splits under $\SO(6)$ into a
label for the singlet and $[IJ]$, so that the only non vanishing
components of $X_{MN}{}^P$ read:
\begin{eqnarray}
X_{I_1J_1,I_2J_2}{}^{I_3J_3}&=&
4\,g\,\delta^{[I_3}_{[I_1}\,\delta_{J_1][I_2}\,\delta_{J_2]}^{J_3]}\,\,,\,\,\,X_{I_1J_1}{}^{I_3J_3}{}_{I_2J_2}=-X_{I_1J_1,I_2J_2}{}^{I_3J_3}\,.
\end{eqnarray}
The tensors $T_{\alpha\beta,AB}{}^{CD}$ and $T_{AB,CD}{}^{EF}$ have
the following general expression:
\begin{eqnarray}
T_{\alpha\beta,AB}{}^{CD}&=&\frac{g}{2}\,f^{I_1
J}{}_{\alpha\beta}\,\left(f^{J J_1}{}_{AB}\,\bar{h}_{I_1J_1}{}^{CD}+
h_{I_1J_1\,AB}\,\bar{f}^{J J_1\,CD}\right)\,,\\
T_{EF,AB}{}^{CD}&=&\frac{g}{2}\,f^{I_1 J}{}_{EF}\,\left(f^{J
J_1}{}_{AB}\,\bar{h}_{I_1J_1}{}^{CD}+ h_{I_1J_1\,AB}\,\bar{f}^{J
J_1\,CD}\right)\,.
\end{eqnarray}
It is useful at this point to use a $\U(6)$ covariant
parametrization of the coset (\ref{scalma}) in which the scalar
fields are described by the tensors
$\phi_{AB},\,\overline{\phi}^{AB}$ in the ${\bf 15}+{\bf
\overline{15}}$. The coset representative will have the following
general form:
\begin{eqnarray}
\mathcal{V}_M{}^{\underline{M}}&=&\mathcal{A}^\dagger\,\exp\left[\left(\begin{matrix}0
& {\bf 0}_{1\times 15} & 0 & \phi_{CD}\cr {\bf 0}_{15\times 1} &
{\bf 0}_{15\times 15} & \phi_{AB} &
\frac{1}{2}\,\bar{\phi}^{EF}\,\epsilon_{EFABCD}\cr
0&\bar{\phi}^{CD}& 0 & {\bf 0}_{1\times 15} \cr
\bar{\phi}^{AB}&\frac{1}{2}\,\phi_{EF}\,\epsilon^{EFABCD} &{\bf
0}_{15\times 1} & {\bf 0}_{15\times 15}
\end{matrix}\right)\right]\,,\nonumber\\
\mathcal{A}&=&\frac{1}{\sqrt{2}}\,\left(\begin{matrix}\bfone
&i\,\bfone\cr \bfone&-i\,\bfone&\end{matrix}\right)\,.
\end{eqnarray}
A bosonic background characterized by the value of the scalar fields
at the origin $\phi_{AB}\equiv 0$ describes  a maximally
supersymmetric (i.e. $\cN=6$) $AdS_4$ background. Indeed, since we
have switched on only a component of the embedding tensor in the
${\bf 21}+{\bf \overline{21}}$, at the origin the T--tensor will lie
in the same representations and thus have vanishing projections on
the ${\bf 35}+{\bf 105}$, which are nothing but the spin $1/2$ shift
matrices $N^A{}_B,\,N^A{}_{BCD}$. On such background then the spin
$1/2$ fields have vanishing supersymmetry variation, while
$S^{AB}=a_1 \,\delta^{AB}$, where $|a_1|=2$. Since
$V_0=V(\phi=0)=-48\,g^2$ one easily  verifies that
$-3\,m_{\frac{3}{2}}^2=-12\,g^2\,|a_1|^2=V_0$, which is condition
(\ref{condvac}) for a maximally supersymmetric $AdS_4$ solution.
Note that the unbroken symmetry  in the vacuum is $\OSp(6/4)\times
\SO(2)$, where the $\SO(2)$ is gauged by the singlet  gauge field
under which no field of the theory is charged, as it should be since
$\SO(2)$ commutes with the supersymmetry generators.
\par Let us analyze the relation between this four
dimensional vacuum solution and the ten dimensional $AdS_4\times
\mathbb{C}P^3$ solution of Type IIA superstring. This higher
dimensional background, as recalled in section \ref{antonio}, is
characterized by a 4- and a 2--form flux $F_{\mu\nu\rho\sigma}=g\,
\epsilon_{\mu\nu\rho\sigma}$, $F_{IJ}=k\, \mathcal{J}_{IJ}$,
$\mathcal{J}_{IJ}$ being the K\"ahler form on $\mathbb{C}P^3$. The
former is invariant under $\SO(6)$ while the choice of the latter
breaks $SO(6)$ into $\U(3)$. We may choose indeed $\mathcal{J}_{IJ}$
to be the $\U(1)$ generator in $\SO(6)$ commuting with $\SU(3)$. The
$\U(4)$-invariant $AdS_4$ vacuum at the origin is likely to describe
this compactification. In fact we may wonder if the flux $F_{IJ}$
enters this effective $\cN=6$ theory as a v.e.v. of a $\U(3)$
invariant scalar field, thus defining a $\U(3)$-invariant vacuum
characterized by two distinct parameters: $g,\,k$. As we shall see
this is not the case. In order to work out all the $\U(3)$ invariant
vacua of the $\cN=6$ supergravity with $SO(6)$ gauging it suffices
to compute the fermion shifts and the scalar potential as a function
of the only complex singlet
$\phi^{sing.}_{AB}=\phi\,\delta^{IJ}_{AB}\,\mathcal{J}_{IJ}$. The
fermion shift tensors read:
\begin{eqnarray}
N^A{}_B&=&-a_2\,
\frac{\phi}{\bar{\phi}}\,\delta^{AD}\,\mathcal{J}_{DB}\,\,,
\,\,\,N^A{}_{BCD}=-3\,a_2\,\delta^{AE}\,\mathcal{J}_{E[B}\mathcal{J}_{CD]}\,\,,\,\,\,
S^{AB}=a_1\,\delta^{AB}\,,\end{eqnarray} where
\begin{eqnarray}
a_2&=&\frac{1}{2}\,e^{-3\,|\phi|}\,(e^{4\,|\phi|}-1)\,\left[\frac{\bar{\phi}}{|\phi|}\,(e^{2\,|\phi|}+1)-(e^{2\,|\phi|}-1)\right]\,,\\
a_1&=&-\frac{1}{4}\,e^{-3\,|\phi|}\,\left[\frac{\bar{\phi}}{|\phi|}\,(e^{2\,|\phi|}+1)^3-(e^{2\,|\phi|}-1)^3\right]\,.\end{eqnarray}
The scalar potential is:
\begin{eqnarray}
V^{(\cN=6)}(\phi,\bar{\phi})&=&-24\,g^2\,e^{-2\,|\phi|}\,(e^{4\,|\phi|}+1)\,.
\end{eqnarray}
From the above result it is clear that the only $\U(3)$ invariant
vacuum of the gauged $\cN=6$ supergravity coincides with the
$\SO(6)$ invariant, maximally supersymmetric, $AdS_4$ background at
the origin. In section \ref{ads4628} we shall show that this $\cN=6$
$AdS_4$ theory does not describe the spontaneously broken phase of a
gauged $\cN=8$ theory, for any gauging. It can be obtained only as a
consistent truncation of the $\SO(8)$-gauged $\cN=8$ theory. The
same holds true for the $\cN=2$ $AdS_4$ theory to be discussed in
next section.
\subsection{$\cN=2$ gauging with $ \SO(2)\times\SO(6)$ gauge group}
As we have seen above, in the absence of hypermultiplets the
$\cN=2$ scalar potential has the general form
\begin{eqnarray}
V&=& -\frac 12 (\Im \cN^{-1})^{\Lambda\Sigma}\cP_\Lambda
\cP_\Sigma + (U^{\Lambda\Sigma} -3 \bar L^\Lambda L^\Sigma)
\cP^x_\Lambda \cP^x_\Sigma\,.
\end{eqnarray}
$\cP^x_\Lambda$ is a constant Fayet--Iliopoulos term, that  in the
gauging at hand  can be chosen as:
\begin{eqnarray}
\cP^x_0=\delta^x_1\,,\qquad \cP^x_\Lambda =0 \mbox{ for } \Lambda \neq 0\,,
\end{eqnarray}
corresponding to the gauging of the global $\SO(2)\subset \SU(2)$ symmetry.
The propotential $\cP_\Lambda$, with
$\cP_{\Lambda =0}=0$ is instead responsible for the gauging of the vector multiplets isometries, along the $G_e=\SO(6)$ Lie algebra.

The $AdS_4$  supersymmetric vacuum corresponds to
\begin{eqnarray}
\cP_\Lambda|_{vac}=0\quad \mbox{ and }\quad U^{00}=0|_{vac}\label{adssusy}
\end{eqnarray}
In the background \eq{adssusy} we then obtain
\begin{eqnarray}
V|_{AdS_4}=-3\,m^2_{3/2}
\end{eqnarray}
where $m_{3/2}= g|L^0|$.

The condition
\begin{eqnarray}
U^{00}=0\,\label{u}
\end{eqnarray}
 which is a necessary condition to preserve supersymmetry,
  is equivalent to set $\mathcal{D}_a L^0= f^0_a=0$ ($a=1,\cdots 15$) on the vacuum.
    Note that \eq{u} is a crucial condition for the gauging. It describes
    how the $\SO(2)$ factor is coupled to the gauge fields. For instance,
    if we would adopt instead a parametrization for the symplectic sections based
    on a cubic prepotential, then we would find $U^{00}=3|L^0|^2$, which corresponds to
    a Minkowski vacuum (rather than anti de Sitter), with broken supersymmetry and flat directions for
    $\cP_\Lambda=0$.
    For a gauge group $G_e$, this would also give solutions with $G_e \to \U(1)^{\mbox{rank }G_e}$ through the Higgs mechanism and would correspond to a no-scale $\cN=2$ supergravity. The standard cubic parametrization corresponds to a manifestly $\SU^*(6)$ invariant setting, since this is the parametrization which comes from dimensional reduction of $D=5$ supergravity. The manifest compact symmetry in this case is $\USp(6)$ rather that $\U(6)$, so the coordinates corresponding to the Cartan decomposition are not special coordinates, which in this setting would correspond to the entry $f^\Lambda_{\underline{\bf 1}}=L^\Lambda$ of the matrix \eq{section}.
     In the Cartan parametrization we have $X^{\bf 15} = f^{\bf 15}_{\underline{\bf 1}}/f^{\bf 1}_{\underline{\bf 1}}$, and the $\SO(6)$ invariant part corresponds to $X^{\bf 15}=0$.

We note that the simplest $\cN=2$ theory which exhibits vacua with
an unbroken $\OSp(2/4)\times G_e$ algebra are $\cN=2$  vector
multiplets minimally coupled to supergravity
\cite{Luciani:1977hp}. In this case one can easily show that the
condition \eq{u} is satisfied in the $G_e$ unbroken phase. These
models, together with their spontaneously broken phases were
studied in \cite{Derendinger:1983rc}. We remark that the special
K\"ahler geometry underlying minimal couplings correspond to the
$\mathbb{C}P^n$ non-compact manifolds $\SU(1,n)/\U(n)$. These are
the only symmetric special geometry which cannot be lifted to five
dimensions.


\subsection{ $\mathcal{N}=6$ and $\mathcal{N}=2$  $AdS_4$ backgrounds from  gauged $\mathcal{N}=8$
theory}\label{ads4628} In this section we show that the $\U(4)$
gauged $\mathcal{N}=2$ and  $\mathcal{N}=6$ theories (the latter
describing the low energy dynamics of Type IIA superstring on a
certain $AdS_4\times \mathbb{C}P^3$ background) cannot be viewed as
spontaneously broken phases of a gauged $\mathcal{N}=8$ theory, they
are instead consistent truncations of the maximal supergravity with
$\SO(8)$ gauging. This implies that the deformation, discussed in
\cite{Nilsson:1984bj,Aharony:2008ug}, which takes $AdS_4\times S^7$
to the $\cN=6$ $AdS_4\times \mathbb{C}P^3$ is not described by the
v.e.v. of a zero-mode on $AdS_4$, i.e. of a scalar field in the
maximal four dimensional model with gauging $\SO(8)$. This is
consistent with the fact that the only $\U(4)$--invariant vacuum
found by Warner in the eighties \cite{warner} has $\cN=0$ and should
correspond to the compactification of $D=11$ supergravity on a
``stretched seven sphere'' discussed in \cite{popewarner}. Here we
shall show, using a group theoretical argument, that no
$\U(4)$-invariant $\cN=6$ vacuum can be found in any gauged $\cN=8$
supergravity.\par We start by noting that in the $\cN=8$ theory,
with respect to the common $\SO(8)$ subgroup of the
$\SL(8,\mathbb{R})$ and $\SU(8)$ symmetry groups, the ${\bf 8}$ of
$\SU(8)$ and the ${\bf 8}$ of $\SL(8,\mathbb{R})$ correspond to the
representations ${\bf 8}_s$ and ${\bf 8}_v$ respectively. The
$\U(4)$ symmetry group of the $\cN=6$ $AdS_4\times \mathbb{C}P^3$
solution, is embedded inside $\SO(8)$ in such a way that the
following branchings hold:
\begin{eqnarray}
{\bf 8}_s&\rightarrow &{\bf 1}_{+1}+{\bf 1}_{-1}+{\bf
6}_{0}\,,\nonumber\\
{\bf 8}_v&\rightarrow &{\bf 4}_{+\frac{1}{2}}+{\bf
\overline{4}}_{-\frac{1}{2}}\,\nonumber\\
{\bf 8}_c&\rightarrow &{\bf 4}_{-\frac{1}{2}}+{\bf
\overline{4}}_{+\frac{1}{2}}\,.\label{8br}
\end{eqnarray}
Consequently the corresponding symmetric tensor product
representations ${\bf 35}_s,\,{\bf 35}_v,\,{\bf 35}_c$ branch in the
following way:
\begin{eqnarray}
{\bf 35}_s&\rightarrow &{\bf 1}_{+2}+{\bf 1}_{0}+{\bf 1}_{-2}+{\bf
6}_{+1}+{\bf 6}_{-1}+{\bf 20}_{0}\,,\nonumber\\
{\bf 35}_v&\rightarrow &{\bf 10}_{+1}+{\bf
\overline{10}}_{-1}+{\bf 15}_{0}\,,\nonumber\\
{\bf 35}_c&\rightarrow &{\bf 10}_{-1}+{\bf
\overline{10}}_{+1}+{\bf 15}_{0}\,,\label{35br}
\end{eqnarray}
the 70 scalar fields transform in the ${\bf 35}_v+{\bf 35}_c$,
which can be described as the self-dual and anti self-dual
components of the 4-times antisymmetric tensor product of the
${\bf 8}_s$,  respectively. We know that the most general gauging
of the $\cN=8$ theory is encoded in an embedding tensor
transforming in the ${\bf 912}$ of $\E_{7(7)}$. This
representation describes not just the plain embedding tensor
$\theta_M{}^n$ defining the gauge algebra, which encodes the
coupling constants of the gauged theory,  but also the T-tensor
$T(\Phi,\theta)$ introduced in (\ref{ttensor0}). Therefore if the
maximal theory with gauge group $\mathcal{G}$ admits a vacuum at
$\langle\Phi\rangle\equiv \Phi_0$ with symmetry group
$\mathcal{G}^\prime\subset \mathcal{G}$, the physical quantities
on such vacuum (masses, couplings etc...) must be defined in terms
of the T-tensor evaluated on this solution, namely
$T_0=T(\Phi_0,\theta)$, which must be a
$\mathcal{G}^\prime$--singlet. Since $T(\Phi,\theta)$ belongs to
the ${\bf 912}$ representation,
 a $\mathcal{G}^\prime$--invariant vacuum is described by a
 $\mathcal{G}^\prime$--singlet ($T_0$)
in the ${\bf 912}$ which  provides the fermion shift tensors
computed on the vacuum. Moreover such quantity is subject to the
quadratic constraints, which amount to the Ward identity on the
fermion shift tensors.
\par With respect to $\SU(8)$ the ${\bf 912}$ branches in the ${\bf
36}+{\bf 420}$, corresponding to the shift tensors $S_{ij}$ and
$N^i{}_{jkl}$ respectively, and the conjugate representations. With
respect to $\SO(8)$ the ${\bf 36}$ branches into ${\bf 1}+{\bf
35}_s$, while the ${\bf 420}$ branches into ${\bf 35}_v+{\bf
35}_c+{\bf 350}$. Therefore the branching of the ${\bf 912}$ with
respect to $\SO(8)$ reads:
\begin{eqnarray}
{\bf 912}&\rightarrow &2\times({\bf 1}+{\bf 35}_s+{\bf 35}_v+{\bf
35}_c+{\bf 350})\,.
\end{eqnarray}
The singlet defines the $\SO(8)$ gauging of de Wit and Nicolai. We
may wonder if the ${\bf 912}$ contains any other singlet, besides
this one, with respect to the $\U(4)$ symmetry of the $\cN=6$
background. Since the ${\bf 350}$ does not contain any $\U(4)$
singlet, from (\ref{35br}) we conclude that the only other singlet
$T_0$ is the one contained in the ${\bf 35}_s$ and corresponds to a
symmetric $8\times 8$ matrix $S^{ij}$ of the form
\begin{eqnarray}
S^{ij}&=&\mbox{diag}(s,\,s,\,s^\prime,\,s^\prime,\,s^\prime,\,s^\prime,\,s^\prime,\,s^\prime)\,.
\end{eqnarray}
So far we have not considered the effect of the quadratic
constraints on the T-tensor $T_0$, which imply the Ward identity
for the fermion shifts. Let us show that a generic component of
T-tensor in the ${\bf 35}_s$ violates the Ward identity, and
therefore does not survive the quadratic constraint. Consider
 a generic $T_0\in {\bf 1}+ {\bf 35}_s$. It can be expressed in terms of a
 symmetric matrix $S^{ij}=S^{ji}$. Since $T_0$ has no
 component in the ${\bf 420}$, it will yield a vanishing dilatino
 shift, $N_i{}^{jkl}=0$, while the gravitino shift will be described by the matrix $S^{ij}$ itself. The Ward
 identity at the origin would read:
 \begin{eqnarray}
V^{(\cN=8)}\,\delta_i^j&\propto& S_{ik}\,S^{kj}\,.
 \end{eqnarray}
 the only solution to the above identity is $S^{ij}\propto
 \delta^{ij}$ ($s=s^\prime$) which corresponds to the $\SO(8)$ gauging $T_0\in {\bf 1}$, with no component in the
 ${\bf 35}_s$. \par
 As far as the $\cN=0$ $\U(4)$--invariant $AdS_4$ studied in  \cite{Nilsson:1984bj,warner,popewarner} is
 concerned, the above argument about the Ward identity does not apply. Indeed the $\SU(4)$ symmetry groups pertaining
to the $\cN=0$ and $\cN=6$ vacua are embedded in inequivalent ways
inside $\SO(8)$, see Appendix \ref{decs}. With respect to the
$\U(4)$ symmetry group of the $\cN=0$ vacuum the following
branching holds:
\begin{eqnarray}
{\bf 8}_c&\rightarrow &{\bf 1}_{+1}+{\bf 1}_{-1}+{\bf
6}_{0}\,,\nonumber\\
{\bf 8}_s&\rightarrow &{\bf 4}_{+\frac{1}{2}}+{\bf
\overline{4}}_{-\frac{1}{2}}\,\nonumber\\
{\bf 8}_v&\rightarrow &{\bf 4}_{-\frac{1}{2}}+{\bf
\overline{4}}_{+\frac{1}{2}}\,.
\end{eqnarray}
Now it is the ${\bf 35}_c$ representation which contains the
$\U(4)$-singlet. A $\U(4)$ invariant T-tensor $T_0$ would then be
a combination of the $\SO(8)$ singlet and the $\U(4)$ singlet in
the ${\bf 35}_c$: $T_0\in {\bf 1}+{\bf 35}_c$. The Ward identity
 would now allow a component of $T_0$ inside ${\bf 35}_c$, since
${\bf 35}_c$ is contained inside the ${\bf 420}$ of $\SU(8)$, and
thus $T_0$ will yield $S^{ij}\propto \delta^{ij}$,
$N^i{}_{jkl}\neq 0$. The singlet $T_0$ in the ${\bf 35}_c$, as any
element of the same representation, can be obtained by acting on
the $\SO(8)$-singlet embedding tensor, defining the $\SO(8)$
gauging, by means of the coset representative $\mathcal{V}$
parametrized by a suitable scalar field $\phi^{ijkl}$, since the
scalar fields transform in the ${\bf 35}_v+{\bf 35}_c$. The v.e.v.
of such scalar field provides the deformation which determines the
$\SO(8)\rightarrow \U(4)$ spontaneous symmetry breaking and the
supersymmetry breaking $\cN=8\rightarrow \cN=0$.

\section{An $\cN=2$ truncation of the $\cN=8$ theory with no vector
multiplets and ten hypermultiplets}\label{nove} We can consider a
different $\cN=2$ truncation of the maximal theory in four
dimensions with with no vector multiplets and ten hypermultiplets.
This is the maximal $\cN=2$ truncation of the $\cN=8$ theory with
no vector multiplets. The scalar fields span the manifold:
\begin{eqnarray}
\mathcal{M}^{(\cN=2)}&=&\frac{\E_{6(+2)}}{\SU(2)\times \SU(6)}\,.
\end{eqnarray}
The global symmetry group of the theory is $G=\U(1)\times
\E_{6(+2)}$, which is a maximal subgroup of $E_{7(7)}$. This
theory can indeed be obtained as a truncation of the four
dimensional maximal supergravity. Since the graviphoton is the
only vector field of the model, we may only gauge one abelian
isometry of the quaternionic manifold. Let us describe all
possible gaugings by means of the embedding tensor. This tensor
belongs to the product of the symplectic representation ${\bf R}$
of the electric and magnetic charges, labelled by $M=1,2$,  and
the adjoint representation of $G$. In this case we have:
\begin{eqnarray}
{\bf R}&=&{\bf 1}_{+3}+{\bf 1}_{-3}\,\,,\,\,\,\,{\bf Adj}(G)={\bf
1}_{0}+{\bf 78}_{0}\,,
\end{eqnarray}
and therefore
\begin{eqnarray}
\theta_M{}^n&\in&{\bf R}\times {\bf Adj}(G)={\bf 1}_{+3}+{\bf
1}_{-3}+{\bf 78}_{+3}+{\bf 78}_{-3}\,.
\end{eqnarray}
The singlets ${\bf 1}_{\pm 3}$ do not correspond to a viable
gauging since they would correspond to gauging the global $\U(1)$
symmetry by means of the graviphoton which is charged itself under
this $\U(1)$. Therefore we are left with
\begin{eqnarray}
\theta_M{}^n&\in&{\bf 78}_{+3}+{\bf 78}_{-3}\,.
\end{eqnarray}
Notice that the above representations enter the branching of the
$\E_{7(7)}$ embedding tensor representation with respect to $G$:
\begin{eqnarray}
{\bf 912}&\rightarrow &{\bf 78}_{+3}+{\bf 78}_{-3}+{\bf
27}_{+1}+\overline{{\bf 27}}_{-1}+{\bf 351}_{+1}+\overline{{\bf
351}}_{-1}\,.
\end{eqnarray}
The fermion fields consist in the gravitini $\psi^\alpha_\mu$,
$\alpha=1,2$, and 20 hyperini $\zeta^{ABC}$, $A=1,\dots, 6$. The
corresponding gauge contribution to the supersymmetry
transformation laws read:
\begin{eqnarray}
\delta\psi^\alpha_\mu
&=&\dots+i\,g\,S^{\alpha\beta}\,\gamma_\mu\,\epsilon_\beta\,,\nonumber\\
\delta\zeta^{ABC} &=&\dots+g\,N_\alpha{}^{ABC}\,\epsilon^\alpha\,.
\end{eqnarray}
The shift tensors $S^{\alpha\beta}$ and $N_\alpha{}^{ABC}$
transform in the representations ${\bf (1,3)}$ and ${\bf (20,2)}$
of the $H=\SU(2)\times \SU(6)$ subgroup of $G$, respectively.
These representations appear, together with their conjugate, in
the branching of the embedding tensor representation with respect
to $H$:
\begin{eqnarray}
{\bf 78}&\rightarrow & {\bf (1,3)}+{\bf (20,2)}+{\bf (35,1)}\,,
\end{eqnarray}
the latter representation correspond to a quantity $N_A{}^B$ which
does not appear in the theory as a fermion shift matrix, though it
enters in the expression of the hyperino mass matrix:
\begin{eqnarray}
M^{ABC,\,EFG}&= &-\frac{1}{24}\,\epsilon^{A_1 A_2 A_3
B[B_1B_2}\,N^{B_3]}{}_{B}\,.
\end{eqnarray}
 If we
interpret this theory as a truncation of the $\cN=8$ one, the
tensor $N_A{}^B$ makes sense as the fermion shift pertaining to
the fermions $\chi^{A\alpha\beta}$ which are truncated.\par If we
denote by $\mathcal{V}$ the coset representative of
$\mathcal{M}^{(\cN=2)}$, the moment maps corresponding to the
gauged $\E_{6(+2)}$ isometry reads:
\begin{eqnarray}
\mathcal{P}^x&\propto &\theta_1{}^n\,\mathcal{V}_n{}^x\,,
\end{eqnarray}
where $\mathcal{V}_n{}^m$ is the matrix representation of
$\mathcal{V}$ in the adjoint representation of $G$. The theory has
an $\cN=2$ $AdS$-vacuum, corresponding to the gauging of a $\U(1)$
inside $\SU(2)$ and zero expectation value of the scalars in the
$H$--covariant parametrization of the coset: $\langle \phi^{\alpha
ABC}\rangle=0$. Indeed such a gauging would correspond to choosing
$\theta\in{\bf (1,3)}$. At the origin the T-tensor coincides with
$\theta$ and thus has zero component on the ${\bf (20,2)}$
representation, implying that $(N_\alpha{}^{ABC})_{\vert vac.}=0$.
This gauging corresponds to a truncation of the $\SO(8)$ gauging
of the $\cN=8$ theory.  The corresponding theory cannot have an
$\cN=2\rightarrow \cN=1$ spontaneous supersymmetry breaking since
it is not coupled to vector multiplets. If we gauge a $\U(1)$
subgroup of $\SU(6)$, $\theta\equiv(\theta_A{}^B)\in {\bf
(35,1)}$. At the origin we would have $(N_\alpha{}^{ABC})_{\vert
vac.}=(S^{\alpha\beta})_{\vert vac.}=0$ which corresponds to an
$\cN=2$ Minkowski vacuum, in which, depending on the eigenvalues
of the $U(1)$ generator $\theta_A{}^B$, a number of
hypermultiplets will become massive.

\section{Acknowledgements}
We thank P. Fr\'e for useful discussions and comments. L. A., R.
D'A., P.A.G. and M.T. wish to acknowledge the Theoretical Physics
Department of  CERN for the kind hospitality. This work is
supported in part by the European Union RTN contract
MRTN-CT-2004-005104, by the Italian Ministry of University (MIUR)
under contracts PRIN 2005-024045 and PRIN 2005-023102. The work of
S.F. is also supported by the D.O.E. grant DE-FG03-91ER40662, Task
C and by the Miller Institute for Basic Research in Science.

\appendix

\section{ Supergroups with zero Killing-Cartan form}
\label{vcf}

We recall the supergroups with zero Killing-Cartan form.
There are three examples
\begin{enumerate}
\item The first example is based on the superalgebra $A(n|n)$ with $n\geq 1$.
The even part of $A(n|n)$ is $A_n \oplus A_n$ and the odd part is
$(n, \bar n) \oplus (\bar n,n)$ where $A_n$ is the usual classical
Lie algebra. The classical real form of this example is
$\mathfrak{psu}(n|n)$ which have subalgebra $\mathfrak{su}(n)
\oplus \mathfrak{su}(n)$, it is generated by supermatrices $2n
\times 2n$ with vanishing supertrace and defined modulo the
identity matrix ${\bf 1}_{2n\times 2n}$ which has vanishing
supertrace. The superalgebra has $\Big(2 n^2-2 \Big| 2\, n^2\Big)$
generators, (it can be shown the corresponding supergroup manifold
has vanishing Ricci curvature).

\item The second example is based on the superalgebra $D(n+1|n)$ with $n\geq1$. The even part
is $D_{n+1} \oplus C_n$ and the odd part is $(2n+2, 2n)$ where
$D_n$ and $C_n$ are the classical Lie algebra series. The real
form is $\mathfrak{osp}(2n +2| 2n)$ (with $n\geq1$) which has the
subalgebra $\mathfrak{so}(2n+2) \times \mathfrak{sp}(2n)$. It is
generated by orthosymplectic supermatrices $4n+2 \times 4n +2$.
The total number of generators is $\Big( 4 n^2 + 4n +1 \Big| 4 n^2
+ 4 n\Big)$, (it can be shown the corresponding supergroup
manifold has vanishing Ricci curvature).

\item The third example is based on the superalgebra $D(2,1; \alpha)$ with $\alpha \not\in \{0,-1\}$.
The even part is $A_1 \oplus A_1 \oplus A_1$ and the odd part is
$(2,2,2)$. The classical real form has bosonic subalgebra
$\mathfrak{sl}(2) \oplus \mathfrak{sl}(2) \oplus
\mathfrak{sl}(2)$. The total number of generators is $(9|8)$.
\end{enumerate}

There are super-cosets with zero Killing forms. They are generated
by symmetric cosets. Here there are some examples:
\begin{enumerate}
\item $\PSU(2,2|4) / \SO(1,4) \times \SO(5)$. This coset has 10 bosonic generators and 32 fermionic
generators. The bosonic subgroup is $\SO(2,4) \times \SO(6) / \SO(1,4) \times \SO(5)$ which is
$AdS_5 \times  S^5$. The fermionic generators are associated to the Killing spinors of the background.

\item $\OSp(6|4) / \U(3) \times \SO(1,3)$. This coset has 10 bosonic generators and 24 fermions.
The bosonic subgroup is $\SO(6) \times \Sp(4) / \U(3) \times
\SO(1,3)$ which is $AdS_4 \times \mathbb{C}P^3$. The fermionic
generators are associated to the Killing spinors which are 24.

\item $\PSU(n+1|n+1) / \SU(n|n+1)$ is also denoted by $\mathbb{C}P^{n|n+1}$ which is a Ricci
flat supermanifold. In the case $n=3$ this is the famous Witten's supertwistor space $\mathbb{C}P^{3|4}$.

\item $\OSp(2n+2|2n)/ \OSp(2n+1|2n)$ is also denoted by $\mathbb{S}^{2n -1| 2n}$ which is the supersphere.

\end{enumerate}

\section{Relevant branchings and decompositions}\label{decs}
\paragraph{$\SO^*(12)\times \SU(2)\subset\E_{7(7)}$}
\begin{eqnarray}
{\bf 56}&\rightarrow & {\bf (12,2)}+{\bf (32,1)}\,,\\
{\bf 133}&\rightarrow & {\bf (1,3)}+{\bf (66,1)}+{\bf (32^\prime,2)}\,,\\
{\bf 912}&\rightarrow & {\bf (12,2)}+{\bf (220,2)}+{\bf (32,3)}+{\bf
(352,1)}\,.
\end{eqnarray}
\paragraph{$\SO^*(12)$ tensor product decompositions}
\begin{eqnarray}
{\bf 32}\times {\bf 66}&\rightarrow & {\bf 32}+{\bf 1728}+{\bf 352} \,,\\
({\bf 32}\times{\bf 32}\times{\bf 32})_s&\rightarrow & {\bf 32}+{\bf
4224}+{\bf 1728}\,.
\end{eqnarray}
\paragraph{$\SU(6)\times \SU(2)\times \U(1)\subset\SU(8)$}
\begin{eqnarray}
{\bf 8}&\rightarrow & {\bf (6,1)}_{+\frac{1}{2}}+{\bf
(1,2)}_{-\frac{3}{2}}\,,\nonumber\\
{\bf 28}&\rightarrow & {\bf (15,1)}_{+1}+{\bf (1,1)}_{-3}+{\bf
(6,2)}_{-1}\,,\nonumber\\ {\bf 36}&\rightarrow &  {\bf
(21,1)}_{+1}+{\bf (6,2)}_{-1}+{\bf
(1,3)}_{-3}\,,\nonumber\\
 {\bf 56}&\rightarrow&{\bf
(20,1)}_{+\frac{3}{2}}+{\bf (6,1)}_{-\frac{5}{2}}+{\bf
(15,2)}_{-\frac{1}{2}}\,,\nonumber\\
{\bf 70}&\rightarrow & {\bf (15,1)}_{-2}+{\bf
(\overline{15},1)}_{+2}+{\bf (20,2)}_{0}\,,\nonumber\\
{\bf 420}&\rightarrow &  {\bf (105,1)}_{+1} +{\bf (20,2)}_{+3}+{\bf
(84,2)}_{-1}+{\bf (15,1)}_{+1}+{\bf (15,3)}_{+1} +{\bf (35,1)}_{-3}
+{\bf (6,2)}_{-1}\,.\nonumber\\&&
\end{eqnarray}
\paragraph{$\SU(6)\times \U(1)\subset\SO^*(12)$}
\begin{eqnarray}
{\bf 12}&\rightarrow & {\bf 6}_{-1}+{\bf \bar{6}}_{+1}\,,\\
{\bf 32}&\rightarrow & {\bf 1}_{+3}+{\bf 15}_{+1}+{\bf \overline{15}}_{-1}+{\bf 1}_{-3}\,,\\
{\bf 32}^\prime&\rightarrow & {\bf 6}_{+2}+{\bf 20}_{0}+{\bf \overline{6}}_{-2}\,,\\
{\bf 66}&\rightarrow & {\bf (35+1)}_{0}+{\bf
\overline{15}}_{+2}+{\bf
15}_{-2}\,,\\
{\bf 352}&\rightarrow & {\bf 35}_{+3}+{\bf (21+15+105)}_{+1}+{\bf
(\overline{21}+\overline{15}+\overline{105})}_{-1}+{\bf
\overline{35}}_{-3} \,,
\end{eqnarray}
\paragraph{$\SU(4)\times \U(1)\subset\SO(8)$.}
There are three
inequivalent $\SU(4)$ subgroups of $\SO(8)$ which we shall denote
here by $\SU(4)_i$, where  $i=v,c,s$. With respect to $\U(4)_i$
the following branchings hold ($i\neq j\neq k\neq i$)
\begin{eqnarray}
{\bf 8}_i&\rightarrow &{\bf 1}_{+1}+{\bf 1}_{-1}+{\bf
6}_{0}\,,\nonumber\\
{\bf 8}_j&\rightarrow &{\bf 4}_{+\frac{1}{2}}+{\bf
\overline{4}}_{-\frac{1}{2}}\,\nonumber\\
{\bf 8}_k&\rightarrow &{\bf 4}_{-\frac{1}{2}}+{\bf
\overline{4}}_{+\frac{1}{2}}\,,\nonumber\\
{\bf 35}_i&\rightarrow &{\bf 1}_{+2}+{\bf 1}_{0}+{\bf 1}_{-2}+{\bf
6}_{+1}+{\bf 6}_{-1}+{\bf 20}_{0}\,,\nonumber\\
{\bf 35}_j&\rightarrow &{\bf 10}_{+1}+{\bf
\overline{10}}_{-1}+{\bf 15}_{0}\,,\nonumber\\
{\bf 35}_k&\rightarrow &{\bf 10}_{-1}+{\bf
\overline{10}}_{+1}+{\bf 15}_{0}\,,\nonumber\\
{\bf 350}&\rightarrow & \overline{{\bf 10}}_{+1}+\overline{{\bf
10}}_{-1}+{\bf 10}_{+1}+{\bf 10}_{-1}+{\bf 6}_{+1}+{\bf
6}_{-1}+{\bf 45}_0+\overline{{\bf 45}}_0+{\bf 64}_{+1}+{\bf
64}_{-1}+{\bf 15}_{+2}+\nonumber\\&&+{\bf 15}_{-2}+2\times{\bf
15}_{0}+{\bf 20}^\prime_0\,,
\end{eqnarray}
the symmetry group associated with the $\cN=6$ vacuum is
$\SU(4)_s\times \U(1)$, while that associated with the $\cN=0$ one
discussed in section \ref{ads4628} is $\SU(4)_c\times \U(1)$.

\section{Fermion mass terms}\label{masst}
In this appendix we write the spin-$1/2$ mass terms for the the
$\cN=6$ and $\cN=2$ truncations of the $\cN=8$ theory. The
spin-$1/2$ mass term for the $\cN=8$ theory reads
\begin{eqnarray}
g\,M^{ijk,\,lmn}\,\overline{\chi}_{ijk}\chi_{lmn}\,,\label{mass8}
\end{eqnarray}
where the mass matrix is expressed uniquely in terms of
$N_i{}^{jkl}$ as follows \cite{de Wit:1982ig}:
\begin{eqnarray}
M^{ijk,\,lmn}&=&-\frac{1}{144}\,\epsilon^{ijkpqr[lm}\,N^{n]}{}_{pqr}\,.
\end{eqnarray}
The above equation allows us to decompose (\ref{mass8}) in terms
of the  $\SU(6)\times\SU(2)\times\U(1)$-irreducible tensors,
introduced in Section \ref{n6n2}, and the spin-$1/2$ fields
pertaining to the $\cN=6$ and $\cN=2$ truncations. The $\cN=6$ and
$\cN=2$ spin-$1/2$ mass terms read:
\begin{eqnarray}
{\cN=6}&:&\nonumber\\&&-\frac{g}{24}\,\epsilon^{A_1A_2A_3C
B_1B_2}\,N^{B_3}{}_C\,\overline{\chi}_{A_1A_2A_3}\chi_{B_1B_2B_3}-\frac{g}{12}\,\epsilon^{BA_1A_2A_3
B_1B_2}\,\oN{}^{B_3}{}_{A_1A_2A_3}\,\overline{\chi}_{B_1B_2B_3}\chi_{B}+\nonumber\\&&+
\frac{g}{16}\,\epsilon^{A_1A_2A_3 B_1B_2
B}\,N^{B_1B_2}\,\overline{\chi}_{A_1A_2A_3}\chi_{B}\,,\\
{\cN=2}&:&\nonumber\\&&\frac{g}{24}\,\epsilon^{\alpha\beta}\,\epsilon^{ABEFGC}\,\oN{}^{D}{}_{EFG}\,
\overline{\chi}_{\alpha AB }\chi_{\beta CD
}-\frac{g}{16}\,\epsilon^{ABCDEF}\,\epsilon^{\alpha\gamma}\,\oN{}^{\beta}{}_{\gamma
EF}\, \overline{\chi}_{\alpha AB }\chi_{\beta CD }\,,
\end{eqnarray}
where the irreducible tensors
$\oN_{D}{}^{ABC},\,N^{AB},\,\oN_{\beta}{}^{\alpha AB}$ were
defined in eq. (\ref{decompN}).

\end{document}